\documentclass[twocolumn,aps,prd,superscriptaddress,nofootinbib]{revtex4-1}

\usepackage{graphicx}
\usepackage{amsmath}
\usepackage{bm}
\usepackage{slashed}
\usepackage{epsfig}
\usepackage{amsfonts}
\usepackage{epstopdf}
\usepackage{hyperref}
\usepackage[subfigure]{graphfig}
\usepackage{braket}
\usepackage{tikz}
\usepackage{hhline}
\usepackage{soul}

\newcommand{\ben}{\begin{eqnarray}}
\newcommand{\een}{\end{eqnarray}}

\newcommand{\bef}{\begin{figure}[!htp]}
\newcommand{\eef}{\end{figure}}

\newcommand{\nn}{\nonumber}

\newcommand{\om}{{\omega}}

\def\be{\begin{equation}}
\def\ee{\end{equation}}
\newcommand{\bea}{\begin{eqnarray}}
\newcommand{\eea}{\end{eqnarray}}

\def\ba{\begin{linenomath*}\begin{equation}}
\def\ea{\end{equation}\end{linenomath*}}

\def\emb{\color{blue}}

\usepackage{soul}

\hypersetup{colorlinks, linkcolor = [rgb]{0,0.0,1.0}, citecolor = [rgb]{0,0.0,1.0}, urlcolor = [rgb]{0,0.0,1.0}}

\begin{document}

\null\hfill\begin{tabular}[t]{l@{}}
  {JLAB-THY-20-3131} 
\end{tabular}

\title{Pion valence quark distribution from current-current correlation in lattice QCD }

\newcommand*{\JLAB}{Thomas Jefferson National Accelerator Facility, Newport News, Virginia 23606, USA}\affiliation{\JLAB}
\newcommand*{\WM}{Physics Department, William and Mary, Williamsburg, Virginia 23187, USA}\affiliation{\WM}
\newcommand*{\CU}{Physics  Department,  Columbia  University,  New  York  City,  New  York  10027,  USA}\affiliation{\CU}
\newcommand*{\PU}{School of Physics and State Key Laboratory of Nuclear Physics and Technology, Peking University, Beijing 100871, China}\affiliation{\PU}
\newcommand*{\CPU}{Center for High Energy physics, Peking University, Beijing100871, China}\affiliation{\CPU}
\newcommand*{\QM}{Collaborative Innovation Center of Quantum Matter, Beijing 100871, China}\affiliation{\QM}

\author{Raza Sabbir Sufian}\affiliation{\JLAB}
\author{Colin Egerer}\affiliation{\WM}
\author{Joseph Karpie}\affiliation{\CU}
\author{Robert G. Edwards}\affiliation{\JLAB}
\author{B\'alint Jo\'o}\affiliation{\JLAB}
\author{Yan-Qing Ma}\affiliation{\PU}\affiliation{\CPU}\affiliation{\QM}
\author{Kostas Orginos}\affiliation{\JLAB}\affiliation{\WM}
\author{Jian-Wei Qiu}\affiliation{\JLAB}
\author{David G. Richards}\affiliation{\JLAB}

\begin{abstract}
We extract the pion valence quark distribution $q^\pi_{\rm v}(x)$ from lattice QCD (LQCD) calculated matrix elements of spacelike  correlations of one vector and one axial vector current  analyzed in terms of QCD collinear factorization, using a new short-distance matching coefficient calculated to one-loop accuracy.  We derive the Ioffe time distribution of the two-current correlations  in the physical limit by investigating the finite lattice spacing, volume, quark mass, and higher-twist dependencies in a simultaneous fit of matrix elements computed on four gauge ensembles. We find  remarkable consistency between our extracted $q^\pi_{\rm v}(x)$ and that obtained from experimental data across the entire $x$ range.  Further, we demonstrate that the one-loop matching coefficient relating the LQCD matrix computed in position space to the $q_{\rm v}^{\pi}(x)$ in momentum space has well-controlled behavior with Ioffe time.  This justifies that LQCD-calculated current-current correlations are good observables for extracting partonic structures by using QCD factorization, which complements to the global effort to extract partonic structure from experimental data.
\end{abstract}

\maketitle
\allowdisplaybreaks


\section{Introduction} 
\label{sec:1}

The pion, being both a Nambu-Goldstone boson and the lightest bound state in quantum chromodynamics (QCD), highlights the challenges in creating consistent theoretical and phenomenological frameworks to describe its partonic structure.  The shape of the pion valence parton distribution functions (PDFs) extracted from experimental data~\cite{Badier:1983mj,Betev:1985pf,Conway:1989fs,Chekanov:2002pf,Aaron:2010ab} in different analyses~\cite{Owens:1984zj,Aurenche:1989sx,Sutton:1991ay,Gluck:1991ey,Wijesooriya:2005ir,Aicher:2010cb,Barry:2018ort} are in sharp contrast among themselves and with perturbative QCD (pQCD)-based frameworks~\cite{Farrar:1979aw,Berger:1979du} at large longitudinal momentum fractions $x$. Central to the disparity is whether the pion PDF has a softer (harder) $(1-x)^2$ $[(1-x)]$ fall-off as $x\to 1$
- various model calculations~\cite{Shigetani:1993dx,Davidson:1994uv,Hecht:2000xa,Chen:2016sno,deTeramond:2018ecg,Bednar:2018mtf,Ding:2019lwe} exemplify this contrast.

The limited available phase space for partonic interactions at large $x$ localizes quantum fluctuations such that large-$x$ dynamics is constrained by confinement, in effect increasing parton correlations as $x\to1$. As the quark distribution at large $x$ is sensitive to nonperturbative quark-gluon dressing, a description of its behavior will also elucidate our understanding of the generation of mass in QCD through dynamical chiral symmetry breaking. Unraveling the complexities of the valence and sea quark contents of the pion is spearheaded by several upcoming experiments - Jefferson Lab tagged deep-inelastic scattering  experiments~\cite{Jlab}, Drell-Yan  measurements at the COMPASS experiment~\cite{Denisov:2018unj} and, also the future Electron-Ion Collider (EIC) facility~\cite{Aguilar:2019teb}. A first-principles lattice QCD (LQCD) determination of the pion valence PDF $q^\pi_{\rm v}(x)$ with controlled statistical and systematic uncertainties is particularly well timed and solicits a synergy of increasing importance between experimental and theoretical efforts.

PDFs are not direct physical observables, such as cross sections, due to the QCD color confinement.
Experimental extraction of $x$-dependent parton physics relies on 
the QCD factorization theorem~\cite{Collins:1989gx} and considerable advancements in global analyses~\cite{Harland-Lang:2014zoa,Dulat:2015mca,Ball:2017nwa,Alekhin:2017kpj,Ethier:2017zbq} of experimental data. 
LQCD cannot calculate PDFs directly due to its Euclidean space formulation.   QCD factorization can, however, connect $x$-dependent parton physics to a class of hadron matrix elements  ``lattice cross sections" (LCSs) -  that are calculable in LQCD and factorizable with perturbative matching. 
Several LQCD methods~\cite{Liu:1993cv,Braun:2007wv,Horsley:2012pz,Ji:2013dva,Ma:2014jla,Radyushkin:2017cyf,Ma:2017pxb} have 
been proposed and developed to probe the light-cone structure of hadrons. 
These approaches have led to significant achievements in recent years, especially in determinations of flavor non-singlet distributions~\cite{Chambers:2017dov,Orginos:2017kos,Alexandrou:2018pbm,Bali:2018spj,Bali:2019ecy,Lin:2018qky,Sufian:2019bol,Izubuchi:2019lyk,Liang:2019frk,Joo:2019jct}.  A proper quantification and mitigation of systematic errors and numerical artifacts present in these calculations and related theoretical challenges still require further insight and development (for a recent review, see~\cite{Cichy:2018mum}).

In this paper, 
we present 
an extraction of the $q^\pi_{\rm v}(x)$ 
from LCSs - LQCD - calculated pion
matrix elements of two local, spacelike-separated and gauge-invariant currents~\cite{Ma:2014jla,Ma:2017pxb}.
These Lorentz covariant matrix elements of two currents spatially separated by a quark propagator are computable on a Euclidean lattice and have a well-defined continuum limit as the lattice spacing $a\to 0$. 
Calculations on four distinct lattice ensembles allows for estimation of systematic errors from finite lattice spacing, volume, and unphysical pion mass extrapolations.  
 From  parity and time-reversal invariance, this vector-axial (VA) current combination is antisymmetric and directly proportional to the $q^\pi_{\rm v}(x)$ with a perturbatively calculable coefficient function that matches this position space LCS to the $q^\pi_{\rm v}(x)$ in momentum space~\cite{Sufian:2019bol}.   With both leading order (LO) and next-to-leading order (NLO) matching coefficients, we extract $q^\pi_{\rm v}(x)$  from the LQCD-calculated pion matrix elements and find that it is remarkably consistent with the same distribution extracted from experimental data  over the entire range of $x$.  We also find that our calculated NLO coefficient function, 
 matching what is calculated in LQCD in position space to PDFs in momentum space,
 is very stable without the large logarithms 
 that are often seen in the perturbatively calculated hard coefficients in momentum space.

 The rest of this paper is organized as follows.  In Sec.~\ref{sec:2}, we first define the pion matrix elements that we calculate in LQCD, and introduce the factorization formalism to match the matrix elements in position space to the PDFs in momentum space.  We then present our perturbative calculation,  and  provide results for the NLO matching coefficients.  We demonstrate the effect of NLO matching coefficient and its perturbative stability in the factorized contribution to the pion matrix elements in position space in Sec.~\ref{sec:3}.  In Sec.~\ref{sec:4}, we explore the stability of the  continuum  limit of  the LQCD-calculated pion matrix elements, and present the numerical extraction of $q^\pi_{\rm v}(x)$.  We then present a discussion of our results in Sec.~\ref{sec:5}, and finally, give our conclusions and outlook in Sec.~\ref{sec:6}.

\section{Calculation of Next-to-leading order perturbative kernel} 
\label{sec:2}

Following our previous work~\cite{Sufian:2019bol}, we consider the following antisymmetrized matrix element in a hadron $h$
\bea \label{eq:matelem}\label{eq:1}
\sigma^{\left[h\right],\mu\nu}_{VA}(\xi,p)&=\xi^4 Z_{V} Z_{A} \bra{h(p)}T\{[\overline{\psi}\gamma^\mu\psi](\xi)\nn \\
      &[\overline{\psi}\gamma^\nu\gamma^5\psi](0)\}\ket{h(p)}+ V\leftrightarrow A,
\eea
where $\sigma^{\left[h\right],\mu\nu}_{VA}$ depends covariantly on the hadron momentum $p$ and spatial separation $\xi$ between the currents; $Z_{V,A}$ are the renormalization constants of the local currents determined in~\cite{Yoon:2016jzj} for the ensembles used in this calculation. A Lorentz decomposition of Eq.~\eqref{eq:matelem} yields  two scalar functions
$T_{i=1,2}^{\left[h\right]}(\omega,\xi^2,p^2)$  where
$\om=-p\cdot\xi$ is the Ioffe time of the process~\cite{Ioffe:1969kf},
 $T^{\left[h\right]}_{i}(\om,\xi^2,p^2)=-T^{\left[h\right]}_{i}(-\om,\xi^2,p^2)$ from parity and time-reversal invariance, and $T^{\left[h\right]}_2$ is power suppressed.
For sufficiently small separations,  
$T_{1}^{\left[h\right]}$, which can be isolated by choosing $\mu=1$ and $\nu=2$, can be factorized~\cite{Ma:2017pxb}
\bea \label{eq:sig}
 T^{\left[h\right]}_{1}(\om,\xi^2,p^2) &=& \sum_{q}\int_0^1dx\,  K(x\om,\xi^2,x^2p^2,\mu^2)\nn \\
&&\times f_{q_{\rm v}/h}(x,\mu^2) + \mathcal{O}(\xi^2\Lambda_{\rm QCD}^2),
\eea
where $f_{q_{{\rm v}/h}}(x,\mu^2)\equiv f_{q/h}(x,\mu^2)-f_{\overline{q}/h}(x,\mu^2)$ are valence PDFs, $\mu^2$ is the factorization scale, and the $K$ is perturbative matching coefficient with $K(x\om,\xi^2,x^2p^2,\mu^2)=-K(-x\om,\xi^2,x^2p^2,\mu^2)$.  Since  $K$ depends on $\xi$, conventional techniques used to calculate matching coefficients in momentum space cannot be applied directly~\cite{NLOcalc}.
To perturbatively calculate  $K(x\om,\xi^2,0,\mu^2)$ with an on-shell struck parton, $k^2=x^2p^2=0$, we 
 could either calculate the matching coefficient directly in position space or
introduce a ``momentum space" matching coefficient with a ``$D$-dimensional" Fourier transform
\bea \label{eq:sigmom}
&&\widetilde{T}_{1}^{\left[h\right]}(\widetilde{\om},q^2) \equiv \int \frac{d^D\xi}{\xi^4} \, e^{iq\cdot\xi}\, T_{1}^{\left[h\right]}(\om,\xi^2,0)\nn \\
&=& \int_0^1 dx\widetilde{K}(x\widetilde{\om},q^2,\mu^2)f_{q_{{\rm v}/h}} (x,\mu^2)+\mathcal{O}(\Lambda_{\rm QCD}^2/q^2),
\eea
where
$D=4-2\epsilon$ and 
$\widetilde{\om}=\frac{2p\cdot q}{-q^2-i0^+}$. 
With the perturbatively calculated $\widetilde{K}$, we can obtain $K$ as
\bea\label{eq:Km}
K(x \om,\xi^2,0,\mu^2)= \xi^4 \int \frac{d^Dq}{(2\pi)^D} e^{-iq\cdot \xi} \widetilde{K}(x\widetilde{\om},q^2,\mu^2)\,.
\eea
To calculate $\widetilde{K}$, we  consider the matrix element of  an on-shell  quark state $q$ in Eq.~\eqref{eq:sigmom}, expand both sides in powers of the strong coupling $\alpha_s$,
and keep up to NLO,
\begin{subequations}\label{eq:matchM}
\bea
\widetilde{T}_{1}^{\left[q\right](0)}(\widetilde{\omega},q^2)=&\int_{0}^1 dx\, \widetilde{K}^{(0)}(x\widetilde{\omega},q^2,\mu^2) f^{(0)}_{q_{\rm v}/q}(x,\mu^2),\nn \\
\\
\widetilde{T}_{1}^{\left[q\right](1)}(\widetilde{\omega},q^2)=&\int_{0}^1 dx\, \widetilde{K}^{(1)}(x\widetilde{\omega},q^2,\mu^2) f^{(0)}_{q_{\rm v}/q}(x,\mu^2)\nn \\
&\hspace{-0.3cm}+\int_{0}^1 dx\,\widetilde{K}^{(0)}(x\widetilde{\omega},q^2,\mu^2) f^{(1)}_{q_{\rm v}/q}(x,\mu^2)\nn \\.
\eea
\end{subequations}
With the well-known $\overline{\text{MS}}$ perturbative PDFs, 
\begin{subequations}\label{eq:fexp}
\begin{align}
f^{(0)}_{q_{\rm v}/q}(x,\mu^2)&=\delta(1-x),\\
f^{(1)}_{q_{\rm v}/q}(x,\mu^2)&= -\frac{1}{\epsilon} \frac{(4\pi)^\epsilon}{\Gamma(1-\epsilon)} \frac{\alpha_s}{2\pi} C_F\left(\frac{1 + x^2}{1 - x}\right)_+ ,
\end{align}
\end{subequations}
$\widetilde{K}^{(0)}$ and $\widetilde{K}^{(1)}$ are 
determined by $\widetilde{T}_{1}^{[q](0)}$ and $\widetilde{T}_{1}^{[q](1)}$ using Eqs.~\eqref{eq:matchM} and \eqref{eq:fexp}. 
The $\widetilde{T}_{1}^{[q](0)}$ and $\widetilde{T}_{1}^{[q](1)}$ are obtained by calculating the two-current (VA) correlator up to $\mathcal{O}(\alpha_s)$ in $D$ dimension.
Due to Ward-Takahashi identities for vector and  axial-vector currents, UV divergences cancel out within one-loop diagrams and we do not need perturbative renormalization, which means $Z_V=Z_A=1$ in the perturbative calculation. One can also verify that  perturbative collinear divergences from $\widetilde{T}_{1}^{[q](1)}$ cancel exactly with $f^{(1)}_{q_{\rm v}/q}$ in Eq.~\eqref{eq:matchM}, resulting in finite $\widetilde{K}^{(0)}$ and $\widetilde{K}^{(1)}$, and thus up to $\mathcal{O}(\alpha_s)$
\begin{align}
&\widetilde{K}(\widetilde{\omega},q^2,\mu^2) =  \bigg\{\frac{1}{1+\widetilde{\omega}}+\frac{\alpha_s C_F}{4\pi } \nn \\ 
&\times\bigg[ \left(\frac{2+2\widetilde{\omega}^2}{\widetilde{\omega}+\widetilde{\omega}^2}\ln(1+\widetilde{\omega})+\frac{3\widetilde{\omega}}{1-\widetilde{\omega}^2}\right) \ln\left(\frac{\mu^2}{-q^2-i0^+}\right) \nn \\
& +\frac{5\widetilde{\omega}}{1-\widetilde{\omega}^2} +\frac{2-2\widetilde{\omega}-\widetilde{\omega}^2}{\widetilde{\omega}+\widetilde{\omega}^2}\ln(1+\widetilde{\omega})  \nn \\
&-\frac{1+\widetilde{\omega}^2}{\widetilde{\omega}+\widetilde{\omega}^2}\ln^2(1+\widetilde{\omega}) \bigg] \bigg\} - (\widetilde{\omega} \to -\widetilde{\omega}).
\end{align}
By performing a Fourier transform, we obtain
\bea \label{eq:FFK}
&&K(\om,\xi^2,\mu^2)= \frac{1}{\pi^2\omega} [ K^{(0)}(\om)+\frac{\alpha_s C_{\rm F}}{2\pi}\nn \\
&& \{K^{(1,0)}(\om)+ K^{(1,1)}(\om) \ln (-\xi^2\mu^2 e^{2\gamma_E}/4)\}],
\eea
with
\bea \label{eq:LO}
 K^{(0)}(\om) = \om \cos\om , 
 \eea 
 \bea\label{eq:NLO}
K^{(1,0)}(\om) &=& \omega \int_0^1 dy \cos(y\omega) \bigg[\frac{1}{2}\delta(1-y) \nn \\
&&-\left(\frac{2\ln(1-y)}{1-y}-\frac{y^2-3y+1}{1-y}\right)_+\bigg]\nn \\
 K^{(1,1)}(\om) &=& -\om \int_0^1 dy \cos(y\om) \bigg(\frac{1+y^2}{1-y}\bigg)_+ \, ,
\eea
where the leading order kernel $K^{(0)}(\om)$ in Eq.~\eqref{eq:LO} is the same as the result in~\cite{Sufian:2019bol}. 
After the integration over $y$, the NLO matching coefficient $K(\omega,\xi^2,\mu^2)$ is very stable and without large logarithms in $\omega$ in the  region  where the lattice QCD data points are available.  
Like the typical perturbatively calculated matching coefficients in momentum space, the NLO matching kernels $K^{(1,0)}(\om)$ and $K^{(1,1)}(\om)$, {\it before the Fourier transform of $y$ into  position space,} have terms with the standard ``+"  prescription  in Eq.~\eqref{eq:NLO}.  The existence of these ``+" prescription  terms is a natural result of perturbative cancellation of infrared (IR)  divergences between the real and virtual contributions (or Feynman diagrams), and these terms have large logarithmic corrections at the point of the IR cancellation~\cite{Sterman:1986aj,Catani:1989ne}.  When these terms are directly convoluted with PDFs in momentum space to derive cross sections, a resummation of such large logarithmic perturbative corrections from the area of IR cancellation is needed to improve the perturbative stability of factorized cross sections so as to be better compared with experimental data near the kinematic threshold~\cite{Aicher:2010cb,Shimizu:2005fp}.   On the other hand, the  QCD factorization proved for the LCSs~\cite{Ma:2017pxb} matches directly the hadron matrix elements calculated in position space to the PDFs in the momentum space, and is valid when the spatial  separation $\xi$ of two currents is sufficiently small $\xi^2\ll 1/\Lambda_{\rm QCD}^2$.
It is this matching of matrix elements in position space to the PDFs in momentum space that helps reduce the perturbative sensitivity to the IR cancelation  that takes  place at a single point in phase space.  As demonstrated in Ref.~\cite{Ma:2017pxb}, the position space matching coefficient $K(\om,\xi^2,\mu^2)$ is perturbatively analytic for all values of $\om$ except $\om\to \infty$.  Technically, the Fourier transform over $y$ in Eq.~\eqref{eq:NLO} 
gives no $\log(\om)$ terms to the $K(\om,\xi^2,\mu^2)$  kernel when $\om$ is in a perturbatively relevant region, and  thus reduces the logarithmic perturbative sensitivity from the terms with the ``+''  prescription.  
With a small  spatial separation between two currents required by the QCD factorization and the limited values of hadron momentum, the relevant $\omega$ is never too large in a practical lattice QCD calculation.

A convergence test of this NLO kernel is demonstrated in Sec.~\ref{sec:3}. 
We highlight
that a large 
hadron momentum $p$ alone does not automatically guarantee QCD factorization of the hadron matrix element in Eq.~\eqref{eq:1} into the PDFs -   and the perturbative kernel, and
contributions from the large $\xi$ region could invalidate the perturbative factorization~\cite{Ma:2017pxb,Sufian:2019bol}. 
It is the smallness of the spatial separation that defines the short-distance probe to see the particle nature of the partons inside a hadron and provides a required hard scale for the QCD factorization.  Although not directly related to this calculation, the need to prove QCD factorization in momentum space is not new and is well known for the transverse momentum $k_T$ part of the transverse momentum dependent (TMD) factorization.   The factorization formalism was proved in its conjugated position $b_T$ space, not in the momentum $k_T$ space~\cite{Collins:1984kg}, and the perturbative matching coefficients, as well as the evolution kernels, are calculated in position space and valid only for small $b_T$.
The perturbative calculation method introduced in this paper can be used not only for the current-current operators, but also for operators defining quasi-PDFs~\cite{Ji:2013dva} and reduced pseudo-ITDs~\cite{Radyushkin:2017cyf} whose factorization to the PDFs are also valid for the region where the spatial separation between two active parton fields is small and much less than $1/\Lambda_{\rm QCD}^2$. More importantly, our method is not restricted to NLO, but can be applied to any perturbative order~\cite{Li:2020xml}.  The main subtlety of the method lies in the Fourier transformation, which must be done in $D$ dimensions as indicated in Eq.~\eqref{eq:Km}. 


\section{Effect of the next-to-leading order kernel on the Ioffe-time distribution} 
\label{sec:3}

\begin{figure}[htp]
		\begin{center}
			\includegraphics[width=3.45in, height=2.35in]{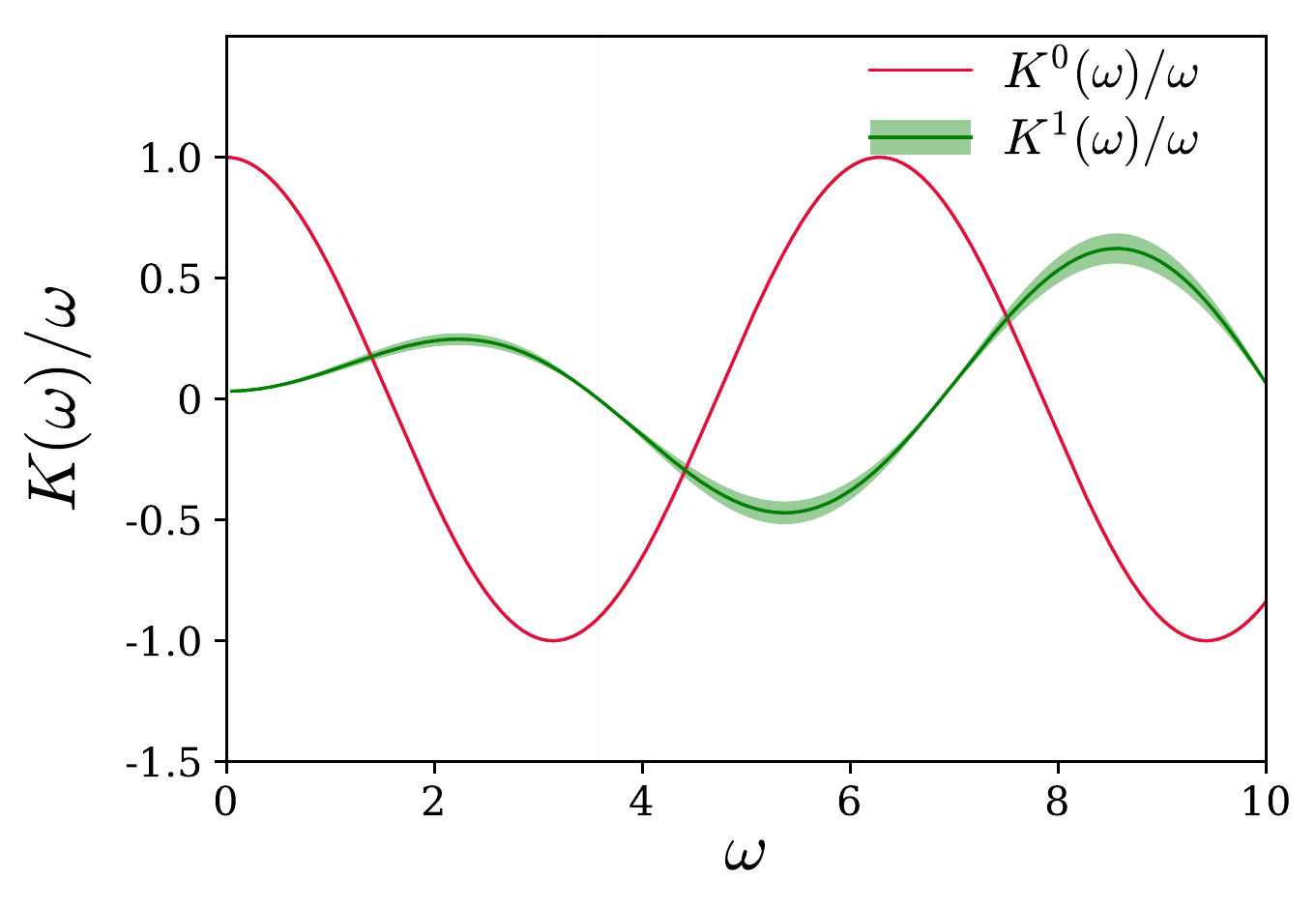}
			\caption{\label{fig:KITD2}
				A comparison between $K^{(0)}(\om)/\om$ and  $K^{(1)}(\om)/\om$ for $\alpha_s(\mu=2\text{ GeV})=0.303$ and $-\xi^2\mu^2=1$. The uncertainty in $K^{(1)}(\om)/\om$ is obtained by a 10\% variation in $\alpha_s$.}
		\end{center}
	\end{figure}

 To demonstrate the effect of the NLO kernel on the Ioffe-time distribution (ITD), we  select  $\alpha_s=0.303$ at $\mu=2$ GeV and $-\xi^2\mu^2=1$  and  compare in Fig.~\ref{fig:KITD2} the $K^{(0)}(\om)/\om$ and $K^{(1)}(\om)/\om$ effects for $\om \neq 0$. The NLO corrections are tiny at small $\om$ and increase very slowly towards large $\omega$; this can be partially understood from the ratio between $K^{(0)}$ and $K^{(1)}$ around $\om=0$:
		\begin{align}
		\frac{K^{(1)}}{K^{(0)}}=\frac{\alpha_s}{3\pi}+\mathcal{O}(\om^2)\approx 0.03+O(\om^2).
		\end{align}
	
\begin{figure}[htp]
		\begin{center}
			\includegraphics[width=3.45in, height=2.35in]{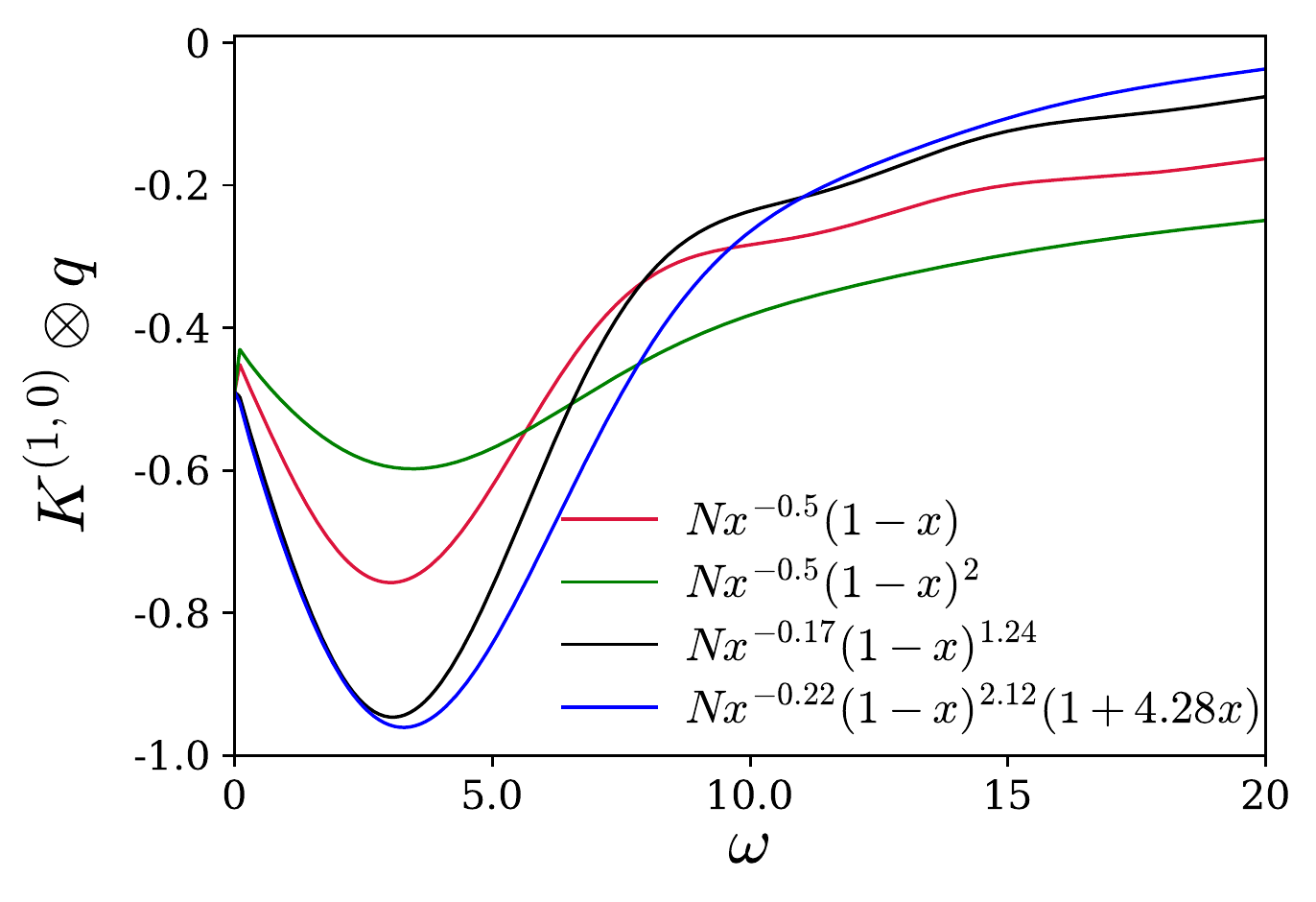}
			\caption{\label{fig:K10q}
				The convolution of the $K^{(1,0)}$ kernel with model PDFs.}
		\end{center}
	\end{figure}

\begin{figure}[htp]
		\begin{center}
			\includegraphics[width=3.45in, height=2.35in]{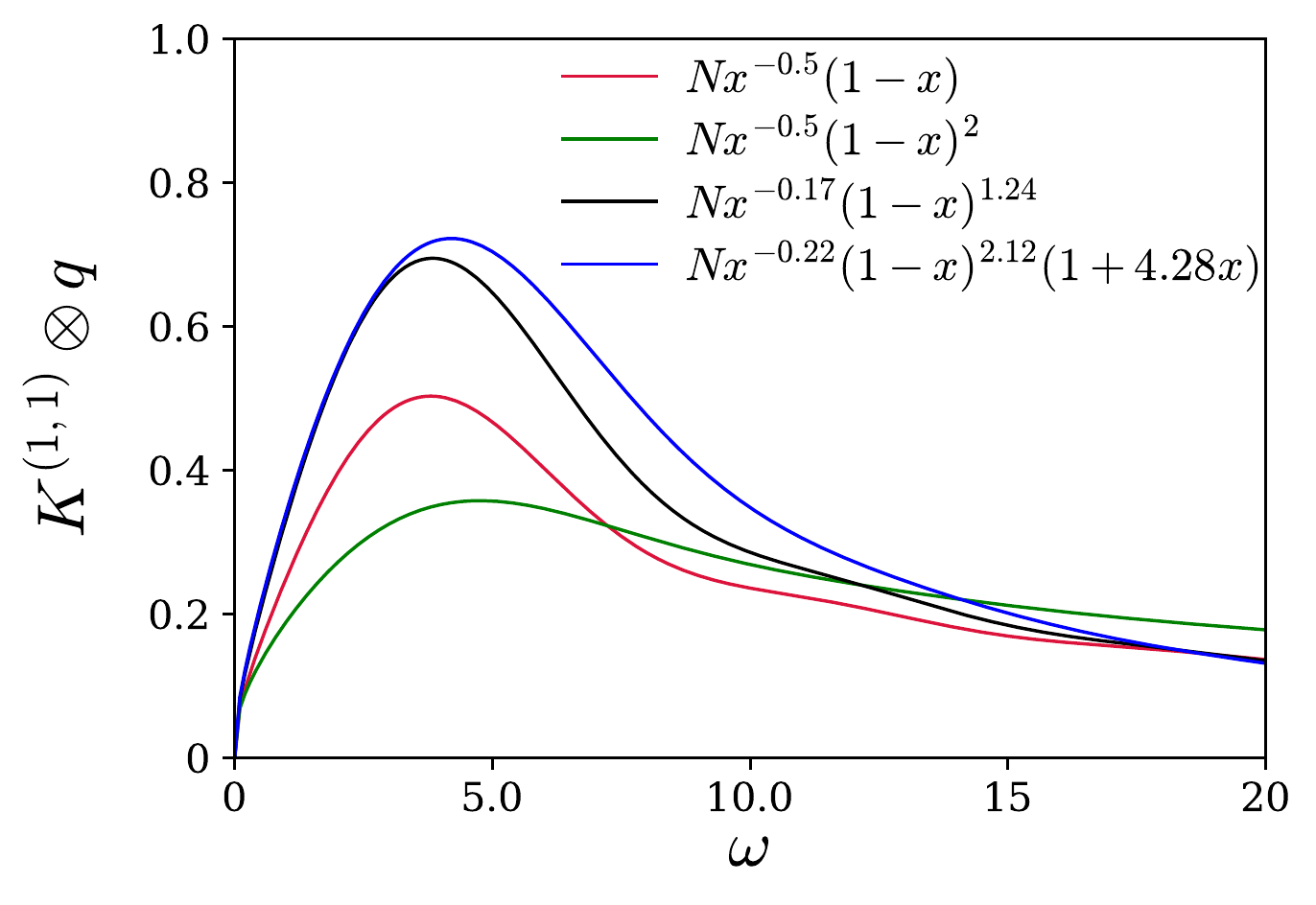}
			\caption{\label{fig:K11q}
				The convolution of the $K^{(1,1)}$ kernel with model PDFs.}
		\end{center}
	\end{figure}

 It is important to note that as an asymptotic series, the relative size of $K^{(1)}/K^{(0)}$ as a function of $\om$ actually diverges as $\om\to\infty$, but, only a small range of $\om$ is relevant for the convolution with PDFs in Eq.~(\ref{eq:1}).  What is important is the size of their convolutions with the PDFs in the relevant Ioffe-time window of the lattice QCD data while keeping  $\xi$ small.  Therefore, it is also useful to demonstrate the effect  the NLO kernels could have with various model PDFs in the Ioffe-time space. The convolutions
\begin{equation}
    K^{(1,i)} \otimes q (\omega) =  \int dx \frac{1}{x \omega} K^{(1,i)}(x\omega) q(x)
\end{equation}
with $i=0,1$ for a few PDFs are shown in Figs.~\ref{fig:K10q} and~\ref{fig:K11q}, respectively. Each of the convolutions have similar features. These convolutions represent the difference between the LCS and the ITD, applying the appropriate factors proportional to $\alpha_s$ and $\ln(-\xi^2\mu^2 e^{2\gamma_E}/4) $. The convolutions all rise to a peak around $\omega\sim 4.0$ and begin to decay to 0. The NLO effects are most significant at the highest Ioffe-time range available to our calculations but the corrections will be smaller for large Ioffe times. These convolutions demonstrate a reassuring feature of the position space matching. These convolutions are at the largest $O(1)$ which means the NLO term will be $O(\alpha_s)$ for the entire region of Ioffe time. These convolutions can be compared with those for matching the reduced pseudo-ITD to the PDF in~\cite{Joo:2019jct}.


\section{Numerical Results \& Extraction of the $q^\pi_{\rm v}(x)$} 
\label{sec:4}
\begin{table}
  \begin{center}
    \setlength\abovecaptionskip{-1pt}
    \setlength{\belowcaptionskip}{-10pt}
    \begin{ruledtabular}
      \begin{tabular}{ccccc}
      ID  & $a$ (fm)  &$m_\pi$ (MeV)  & $L^3\times N_t$ & $N_{\rm cfg}$\\
      \hline
      $a127m413$ & $0.127(2)$ & 413({4}) & $24^3\times 64$ & 2124 \\
      \hline
      $a127m413L$ & $0.127(2)$ & 413({5}) & $32^3\times 96$ & 490 \\
      \hline
      $a94m358$ & $0.094(1)$ & 358({3}) & $32^3\times 64$ & 417 \\
      \hline
      $a94m278$ & $0.094(1)$ & 278(4) & $32^3\times 64$ & 503 \\
      \end{tabular}
    \end{ruledtabular}
    \caption{ \label{tab:lat} Parameters for each gauge ensemble used in this work: lattice spacing, pion mass, spatial and temporal sizes, and number of configurations used.}
  \end{center}
\end{table}
\addtolength{\textfloatsep}{-0.3in}

 The LQCD calculation of  the pion matrix element in Eq.~(\ref{eq:1})
is carried out on four different 2+1 flavor QCD ensembles (listed in Table~\ref{tab:lat}) using the isotropic-clover fermion action generated by the JLab/W\&M Collaboration~\cite{lattices}. 
 We refer  to~\cite{Sufian:2019bol} for details about the implementation of a modified sequential source technique, and a combination of Jacobi and momentum smearing to obtain matrix elements for a given momentum $p$ and spatial separation $\xi$ between the currents. In this calculation of the forward matrix elements, the pion source-sink separation $T$ is systematically increased, while holding fixed the current insertion time $t=T/2$, ensuring identical excited-state contamination from both source and sink sides. To extract the desired matrix elements, we assume the following forms of two- and four-point correlation functions:
\bea
C_{2pt}(T) &=& A\, e^{-m_0 T}\nn \\
C_{4pt}(T) &=& e^{-m_0 T} (B+D\, e^{-\Delta m T}),
\eea
and perform simultaneous correlated fits to the two- and four-point functions. We verify that the value of the ground-state energy $m_0$ obtained from this simultaneous fit is consistent with that obtained from $C_{2pt}(T)$ alone and also agrees with the energy-momentum dispersion relation. 

\begin{figure}[htp]
\begin{center}
\includegraphics[width=3.4in, height=4.6in]{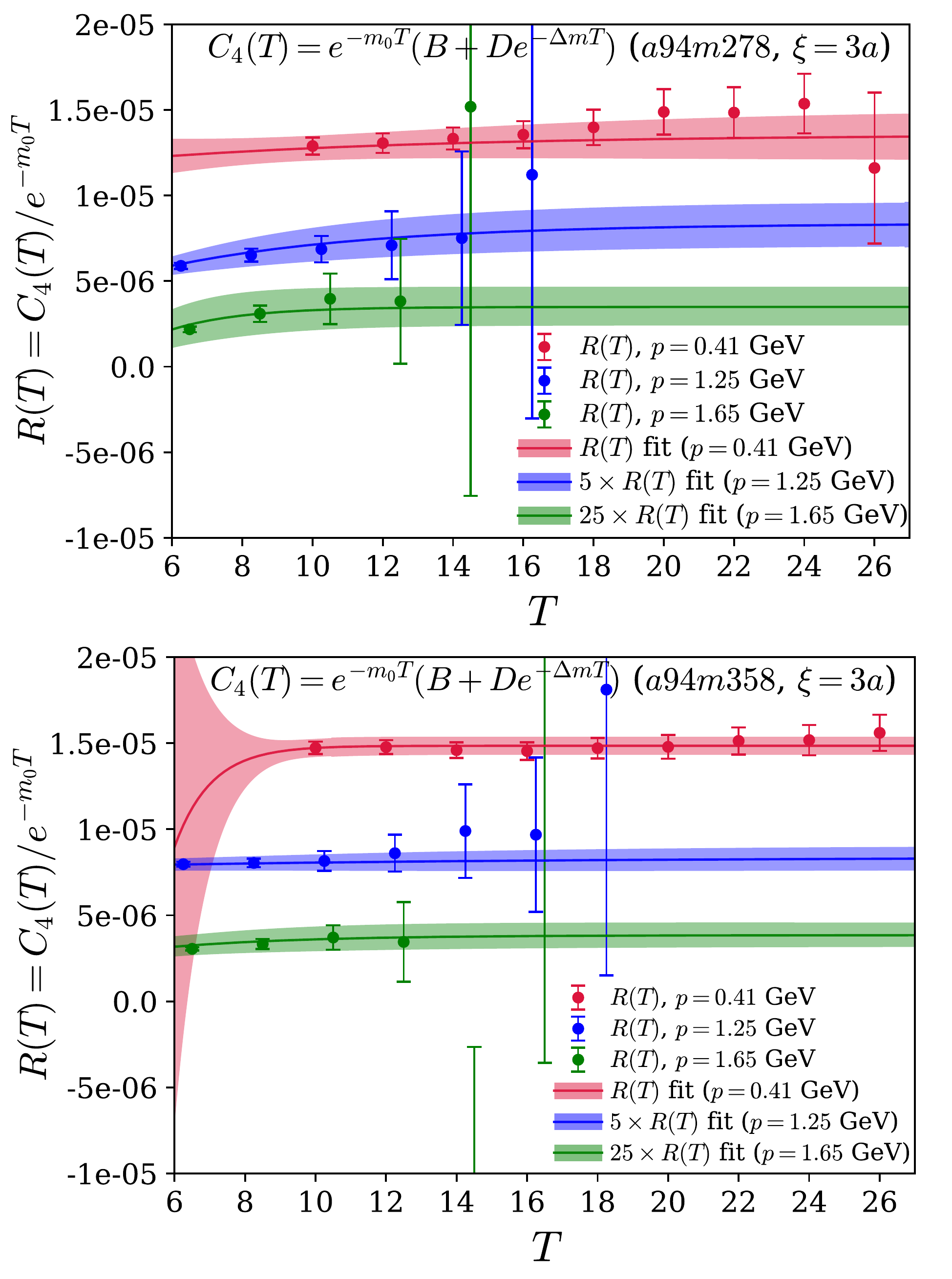}
\caption{\label{fig:matelemes}
Removal of the leading ground-state time dependence exposes the desired matrix elements in the large $T$ limit, shown here for ensembles $a94m278$ (above) and $a94m358$ (below) for current separations $\xi = 3a$. High momenta data rescaled for $S/N$ comparison. }
\end{center}
\end{figure}

In Fig.~\ref{fig:matelemes}, we present fit results of the ratio $C_{4pt}(T)/e^{-m_0 T}$ on the ensembles $a94m278$ and $a94m358$ for momenta  in the range $p\in\lbrace0.41-1.65\rbrace$ GeV and current separation $\xi=3a$, both $p$ and $\xi$ in along the $z$ direction, to demonstrate how reliably we can extract the asymptotic value of $B$, and hence the ITD from $B/A$. The numerical challenges manifest in this formalism are reflected in the signal-to-noise ratio ($S/N$) of the largest momentum $p=1.65$ GeV relative to that of the smallest $p=0.41$ GeV; the former is nearly 3 times smaller. Despite this, we can fit these data up to at least $T=14(\sim1.32\text{ fm})$ even for the largest momentum $p=1.65$ GeV on the lightest pion mass $m_\pi=278$ MeV ensemble. In all the fits, we use the time window such that $S/N \geq 1$.   The Wilson clover fermion action explicitly connects adjacent lattice sites, introducing spurious contact terms in the $\xi=a$ matrix element signals. These data are consequently neglected from our analysis.
\addtolength{\textfloatsep}{-0.10in}

\begin{figure}[htp]
\begin{center}
\includegraphics[width=3.45in, height=2.35in]{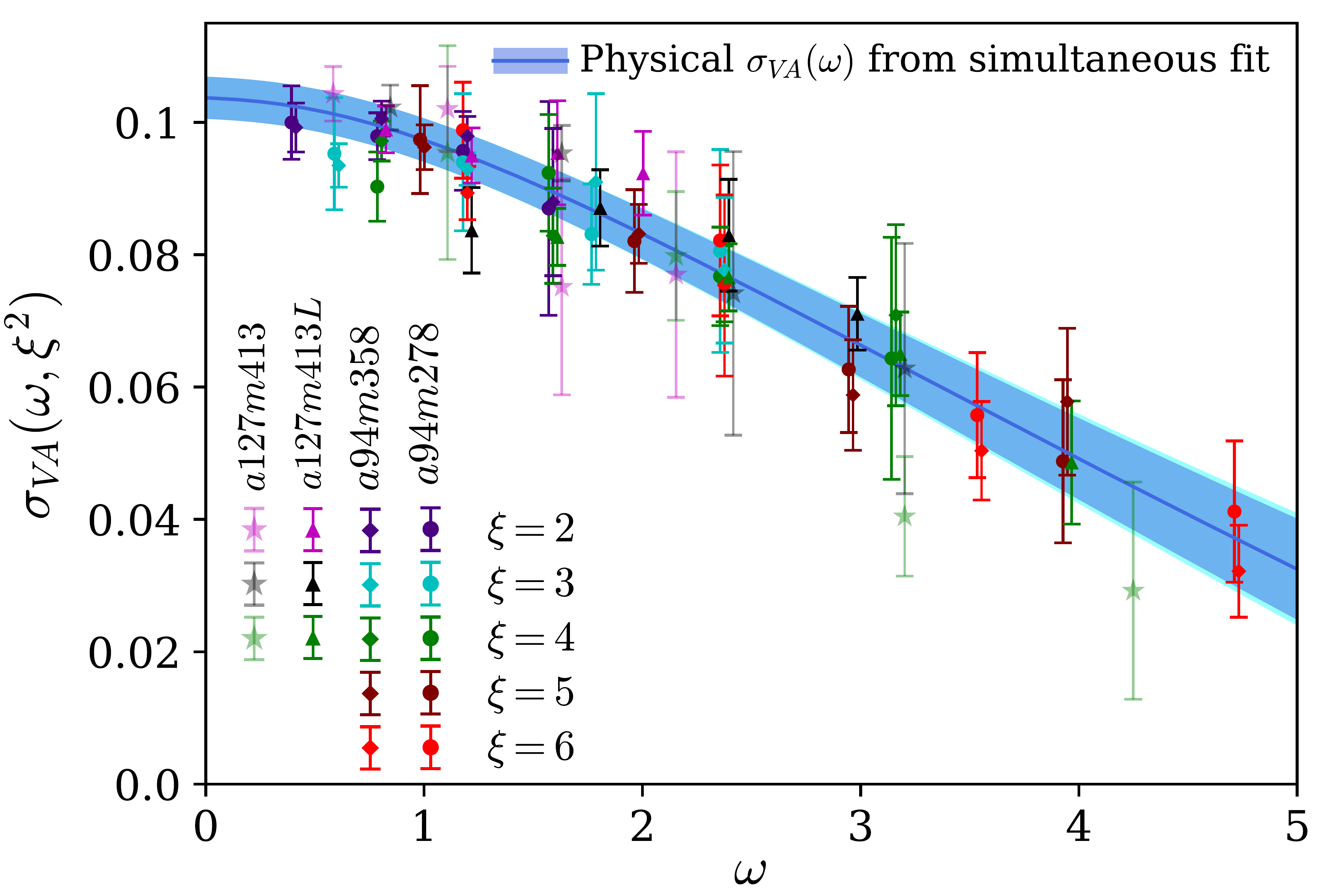}
\caption{\label{fig:ITD}
Simultaneous fit to the antisymmetric VA current matrix elements on four different ensembles. The blue band  indicates the ITD in the physical limit. The outer cyan band shows the combined statistical and systematic uncertainties of fit~\eqref{zfit} added in quadrature.}
\end{center}
\end{figure}

The matrix elements computed across the four gauge ensembles are shown in Fig.~\ref{fig:ITD}. We only include $\lvert\xi\rvert\le 0.56$ fm in our analysis so that $\xi$ is sufficiently smaller than $\Lambda_{\rm QCD}^{-1}$, thereby ensuring the validity of the short-distance factorization and minimizing higher-twist contributions from large $\xi$.
Exploiting the analyticity of the LCS $T^{\pi}_{1}(\om,\xi^2)= \sigma^{12}_{\rm VA}(\om,\xi^2)$ in $\om$ and denoting $\sigma^{12}_{\rm VA}(\om,\xi^2)\equiv \sigma_{\rm VA}(\om,\xi^2)$ in the rest of the article and figures, we obtain the functionally unknown ITD using a flexible $z$-expansion fit~\cite{Boyd:1994tt,Bourrely:2008za} supplemented with chiral, continuum, finite volume~\cite{Briceno:2018lfj} and higher-twist corrections:
\bea \label{zfit}
\sigma_{\rm VA}(\om,\xi^2) &=& \sum_{k=0}^{k_{\rm max}=4} \lambda_k \tau^k +b_1 (m_\pi-m_{\pi,{\rm physical}})+b_2 a \nn \\
&+& b_3 \xi^2 + b_4 a^2p^2 + b_5 e^{-m_\pi(L-\xi)}\, ,
\eea
\bea
\!\!\!\!\!\!\!\!\!\!\!\!\!\!\!\!\!\!\!\!\!\!\!\!\!\!\!\!\!\!\!\!\!\!\!\!\!\!\!\!\!\!\!\!\!\!\!\!\!\!\!\!\!\!\!\!\!\!\!\!{\rm where} \quad \tau = \frac{\sqrt{\om_{\rm cut}+\om}-\sqrt{\om_{\rm cut}}}{\sqrt{\om_{\rm cut}+\om}+\sqrt{\om_{\rm cut}}}
\eea
and $m_{\pi,{\rm physical}} \simeq 0.14$ GeV is the physical pion mass. Higher-order terms $\left(k_{\rm max}>4\right)$ have no statistical significance and are not considered.  
 
  The correction terms of  Eq.~\eqref{zfit} are selected based on each being the dominant   contribution of its type. We consider  now different possible correction terms in  the fit   to Eq.~\eqref{zfit}, such as $a^2$, $m^2_\pi$,  $Le^{-m_\pi(L-\xi)}$, $\sqrt{L}e^{-m_\pi(L-\xi)}$. These corrections are presented in Table~\ref{tab:systematics}. The second column indicates the value of the fitted coefficient of the correction terms and the $\lambda_k$ columns indicate the effect of these corrections on the $z$-expansion fit parameters used to obtain the physical limit $\sigma_{\rm VA}(\omega)$ distribution. We note that for all additionally considered corrections,  the effect is indeed observed to be less than the original corrections of Eq.~\eqref{zfit} and the determination of $q^\pi_{\rm v}(x)$ remains unaffected.
\begin{table*}[htp]
  \centering
  \begin{tabular}{|c|c|c|c|c|c|c|c|}
  \hline
    Correction term  & Fit coefficient & $\lambda_0$   &$\lambda_1$  &  $\lambda_2$ & $\lambda_3$ & $\lambda_4$ & $\chi^2/{\rm d.o.f.}$\\
    \hhline{|=|=|=|=|=|=|=|=|}
    $a^2$ & $-0.049(34)$ &  $0.0104(3)$ & $-0.006(3)$ & $-0.028(9)$ & $-0.901(391)$ & $0.124(135)$ & 1.26 \\
    \hline
    $(m_\pi^2-m^2_{\pi,{\rm physical}})$ & $0.15(12)$ & $0.0104(3)$ & $-0.006(3)$ & $-0.029(10)$ & $-0.926(388)$ & $0.118(132)$ & $1.18$\\
    \hline
    $Le^{-m_\pi(L-\xi)}$ & $0.007(3)$ & $0.0104(3)$ & $-0.006(3)$ & $-0.028(10)$ & $-0.915(402)$ & $0.121(136)$  & 1.22\\
    \hline
    $\sqrt{L}e^{-m_\pi(L-\xi)}$ & $0.026(14)$ & $0.0104(3)$ & $-0.006(3)$ & $-0.029(10)$ & $-0.914(403)$ & $0.121(136)$ & 1.21 \\
    \hhline{|=|=|=|=|=|=|=|=|}
  \end{tabular}
  \caption{Fit parameters of different correction terms in fit Eq.~\eqref{zfit} for the investigation of systematic uncertainties in $\sigma_{\rm VA}(\omega)$.}
  \label{tab:systematics}
\end{table*}

In addition, as seen in Fig.~\ref{fig:ITD} the effects of other possible correction terms such as $m^2_\pi$, $a^2$, $L e^{-m_\pi(L-\xi)}$, $\sqrt{L} e^{-m_\pi(L-\xi)}$ are observed to be very mild. We choose $\om_{\rm cut}=1.0$ as used in~\cite{Joo:2019bzr}; other choices of $\om_{\rm cut}$ were observed to have no effect on the final band in the physical limit and vanishing higher-twist $\mathcal{O}(\xi^2)$  contributions. The blue band in Fig.~\ref{fig:ITD} shows such $\sigma_{\rm VA}(\om)$ distribution after $b_i$ corrections in Eq.~\eqref{zfit} are subtracted, and where the error band is determined from the $\lambda_k$ covariances. The fit yields 
 \bea
 &&\lambda_0=0.104(3), \, \lambda_1=-0.006(3), \, \lambda_2=-0.029(9), \nn \\
 && \lambda_3=-0.907(404),\, \lambda_4=0.124(136),   \nn \\
 && b_1=0.174(96), \, b_2=-0.083(43), \, b_3=-0.0004(7),   \nn \\
 && b_4=0.007(8), \, b_5=0.102(51)
 \eea
  with $\chi^2/{\rm d.o.f}=1.20$.  As can be seen in Fig.~\ref{fig:ITD}, there appears to be completely negligible $\xi$ effects either higher twist or DGLAP~\cite{Gribov:1972ri,Altarelli:1977zs,Dokshitzer:1977sg}. Therefore, we will assign the $\xi=2 \times 0.094$ fm, the shortest $\xi$ used in this study in the factorization formula~\eqref{eq:FFK} while matching the position space LCS to $q^\pi_{\rm v}(x)$ distribution.  With the physical $\sigma_{\rm VA}(\om)$ distribution in hand, we can immediately extract the physical $q^\pi_{\rm v}(x)$ with no further extrapolations. 
\begin{figure}[htp]
\begin{center}
\setlength\abovecaptionskip{-2pt}
\setlength\belowcaptionskip{-6pt}
\includegraphics[width=3.45in, height=2.5in]{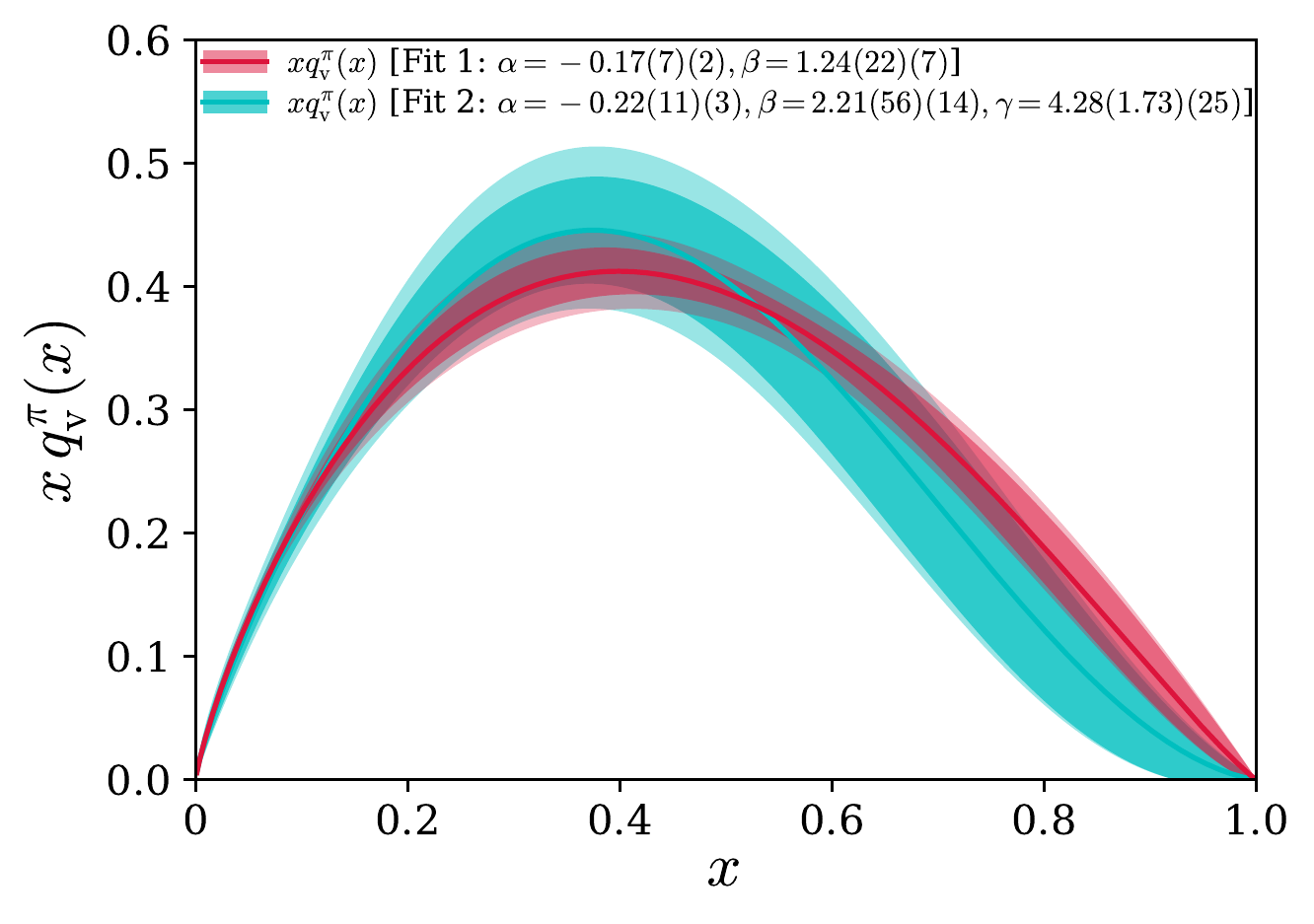}
\caption{\label{fig:xPDF}
The  pion  valence  quark distribution  obtained  from fitting the convolution of $q^\pi_{\rm v}(x)$ and the NLO perturbative kernel~\eqref{eq:FFK} to the determined $\sigma_{\rm VA}(\om)$ distribution in the fit Eq.~\eqref{zfit}. Fits 1 and 2 label the 2- and 3-parameter functional forms in Eq.~\eqref{eq:qxform}.  }
\end{center}
\end{figure}
%

 The extraction of $q^\pi_{\rm v}(x)$ is achieved  by  numerically evaluating the convolution of the NLO kernel equation~\eqref{eq:FFK} and the following phenomenologically motivated functional forms of the PDF:
\bea \label{eq:qxform}
q^\pi_{\rm v}(x) = \frac{x^\alpha(1-x)^\beta(1+\gamma x)}{B(\alpha+1,\beta+1)+ \gamma B (\alpha+2,\beta+1)}
\eea
using the library ROOT~\cite{ROOT}. The high correlation of the sampled $\sigma_{VA}(\omega)$ data guarantees that increasing the sampling density or varying the number of derived pseudodata samples will have no impact on the $q^\pi_{\rm v}(x)$ fit parameters. 

 The parameters in the PDF parametrization equation~\eqref{eq:qxform} are determined by fitting the convolution of the model PDF and the NLO perturbative kernel to $\sigma_{\rm V A}(\omega)$ in a manner similar to Ref.~\cite{Joo:2019bzr}, where the Ioffe-time zero point is fixed in this calculation by the LO$+$NLO perturbative kernel.  The isolation of $\sigma_{VA}(\omega)$ is a multistep process, and begins by performing a correlated fit of lattice data from all four ensembles according to Eq.~\eqref{zfit}. This yields $\sigma_{\rm V A}(\omega,\xi^2)$ plus corrections. Removing the $b_i$ corrections from the obtained $\sigma_{\rm VA}(\omega,\xi^2)$ distribution, we obtain the blue band indicated by $\sigma_{\rm V A}(\omega)$ - now in the physical limit.  The covariance matrix of the $\lambda_k$ coefficients from the correlated $z$-expansion fit provides an error estimate of the $\sigma_{\rm VA}(\omega)$ physical distribution. We choose 30 correlated data points from the continuum band of $\sigma_{V A}(\omega)$, equally spaced in the Ioffe-time interval $\omega\in [0-4.71]$; a number in accordance with the 20 data points available from the $a94m278$ and $a94m358$ lattice ensembles. Using the mean and covariance matrix of these data points, we create 200  Gaussian distributed pseudodata samples with appropriate correlations and perform the following numerical fit 
\bea \label{eq:numfit}
\sigma_{V A}(\omega) = \int_0^1 dx\, K^{\rm LO + NLO}(x,\omega)\, q^\pi_{\rm v} (x)
\eea
to obtain $q^\pi_{\rm v} (x)$. As these discrete values resulting from the fit in Eq.~\eqref{zfit} are highly correlated, the addition of more discrete data points from the fitted $\sigma_{\rm VA}(\omega)$ distribution  does not improve the outcome of the $q^\pi_{\rm v} (x)$ fit parameters. We confirmed this by increasing the number of $\sigma_{\rm VA}(\omega)$ sampling points to 100. A similar result is obtained if one chooses 20 sampling points or less, as was done in our previous work (Ref.~\cite{Sufian:2019bol}). One can also see that increasing or decreasing the number of pseudodata samples from 200 will  not have any impact on the $q^\pi_{\rm v} (x)$ fit parameters, again due to the $\sigma_{\rm V A}(\omega)$ data correlations. What is required to improve the $q^\pi_{\rm v} (x)$ fit parameters is a larger range of Ioffe time.  In the above fit, we have used the constraints $\alpha \leq 0$ and $\beta \leq 4$.

For the above fit, we use $\alpha_s=0.303$ at the initial scale $\mu_0=2\text{ GeV}$~\cite{Buckley:2014ana}. Systematic uncertainties in each PDF parameter set are estimated by a 10\% variation in $\alpha_s$ as in ~\cite{Joo:2019bzr}. The 2-parameter fit, by fixing $\gamma=0$ in Eq.~\eqref{eq:qxform}, yields 
\bea \label{eq:2params}
\alpha = -0.17(7)_{\rm stat}(2)_{\rm sys}, \quad \beta = 1.24(22)_{\rm stat}(7)_{\rm sys}
\eea
with $\chi^2/{\rm d.o.f}=1.41$. Stated uncertainties are statistical (systematic) first (second). In a 3-parameter fit, with an unconstrained $\gamma$, we obtain
\bea \label{eq:3params}
\alpha&=&-0.22(11)_{\rm stat}(3)_{\rm sys}, \, \beta = 2.12(56)_{\rm stat}(14)_{\rm sys},\nn \\
  \gamma &=& 4.28(1.73)_{\rm stat}(25)_{\rm sys}
\eea
with $\chi^2/{\rm d.o.f} \approx 1.29$.  The present calculation has achieved a better statistical precision in the $\beta$ value compared to the previous LCS determination~\cite{Sufian:2019bol} where it was found in a 3-parameter fit $\beta=1.93(68)$. Inclusion of an additional $\rho\sqrt{x}$-term in~\eqref{eq:qxform} was found to be consistent with zero. Commensurate $\chi^2/{\rm d.o.f}$ between fits~\eqref{eq:2params} and~\eqref{eq:3params} limits the selection of one fit over another based solely on the goodness of the fit. These fits are shown in Fig.~\ref{fig:xPDF}. We elected not to extrapolate our ITD obtained from our $z$-expansion fit beyond the largest Ioffe time $\om=4.71$ when determining the PDF. It has been shown~\cite{Karpie:2019eiq} when using sophisticated inversion methods  that the large-$x$ behavior is well reproduced even with the limited range in Ioffe time.


\section{Discussion}  
\label{sec:5}

\begin{figure}[htp]
\begin{center}
\includegraphics[width=3.45in, height=2.45in]{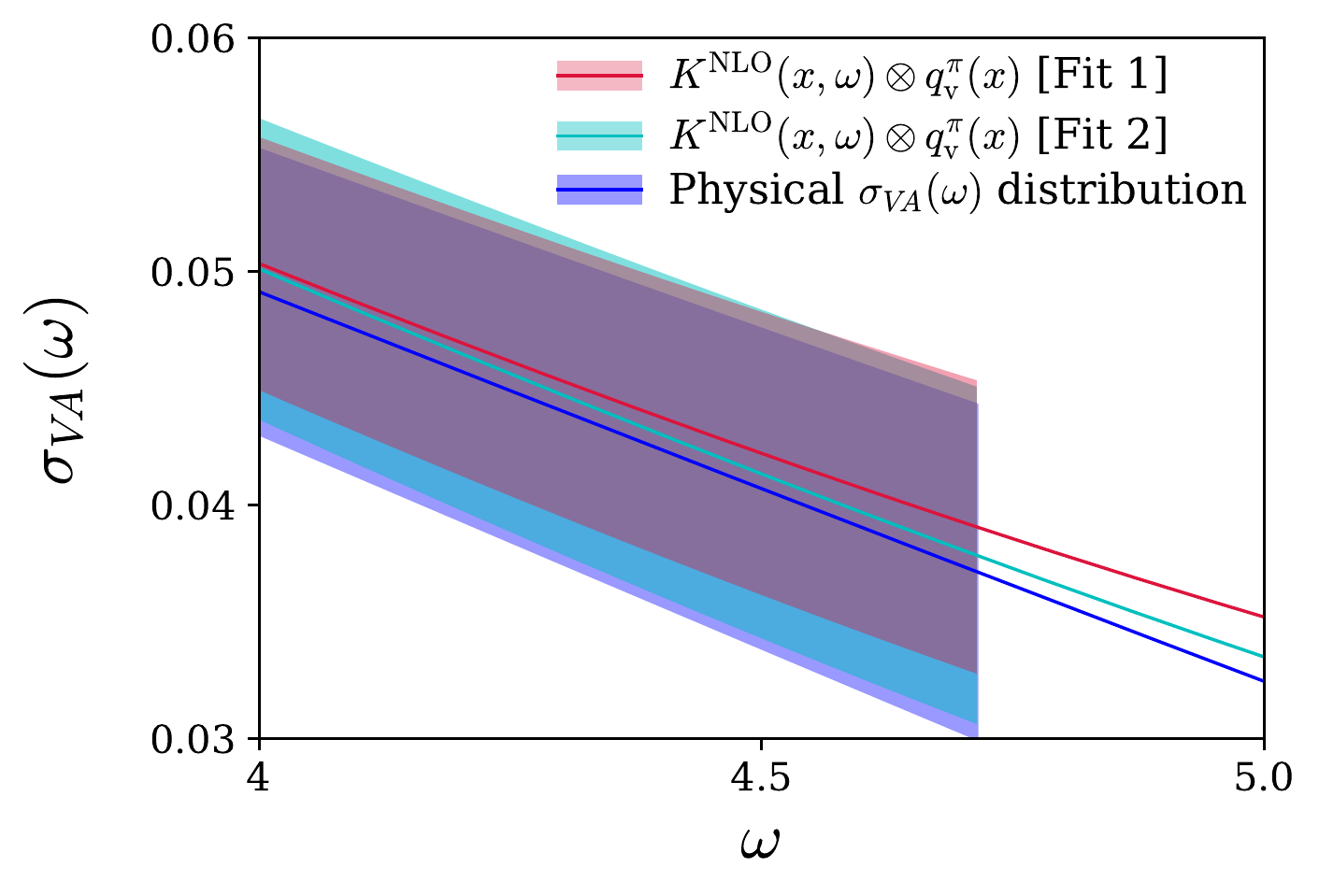}
\caption{\label{fig:ITDrecons}A comparison of the reconstructed $\sigma_{\rm VA}(\om)$ distribution using Eq.~\eqref{eq:numfit} for $4.0<\om<5.0$ from the PDF fits and that obtained from~\eqref{zfit}. Fits 1 and 2 label the 2- and 3-parameter functional forms in Eq.~\eqref{eq:qxform}.  }
\end{center}
\end{figure}

As shown in Fig.~\ref{fig:ITDrecons}, extrapolating the central value of the $\sigma_{\rm VA}(\om)$ distribution from the $z$-expansion fit (blue) and the associated 2- (red) and 3-parameter (cyan) fits reveals that precise LQCD data at large $\om$ are required to distinguish between different large-$x$ behaviors of $q^\pi_{\rm v}(x)$.  
We validate our PDF fitting procedure by reconstructing the $\sigma_{\rm VA}(\omega)$ distribution by convolving the NLO kernel with the PDFs obtained from the pseudodata samples. The $\sigma_{\rm VA}(\om)$ distribution reconstructed from the 2-parameter fit underestimates the uncertainty of the distribution in the physical limit by about 8\%-12\% for $\om >4$, and starts to deviate from the blue band as $\om$ increases. This observation together with the smaller $\chi^2/\text{d.o.f}$ favors the $q^\pi_{\rm v}(x)$ extracted using the 3-parameter fit~\eqref{eq:3params}. For a fixed $\alpha$, one can show that the ITD falls off faster for a smaller $\beta$ as a function of $\om$ compared to that for a larger $\beta$ in a 2 or 3 or more parameter PDF functional form. Therefore, precise data at higher Ioffe time $(\om \sim 8-10)$ will provide  
a better 
discrimination between different $\beta$ values in a future LCS calculation.

While PDFs can minimally be described by the $x^\alpha (1-x)^\beta$ functional form, encompassing the Regge theory~\cite{Regge:1959mz} and pQCD based power counting rules~\cite{Brodsky:1973kr}, modern global analyses~\cite{Harland-Lang:2014zoa,Dulat:2015mca,Ball:2017nwa,Ball:2017nwa} inform our decision to allow for an interpolating function between these small-$x$ and large-$x$ regions and thus a better and less biased description of PDFs. In particular, the fit~\eqref{eq:3params} includes the possibility of $\gamma=0$ and is more flexible.

\begin{figure}[htp]
\begin{center}
\setlength\abovecaptionskip{-2pt}
\setlength\belowcaptionskip{-2pt}
\includegraphics[width=3.45in, height=2.45in]{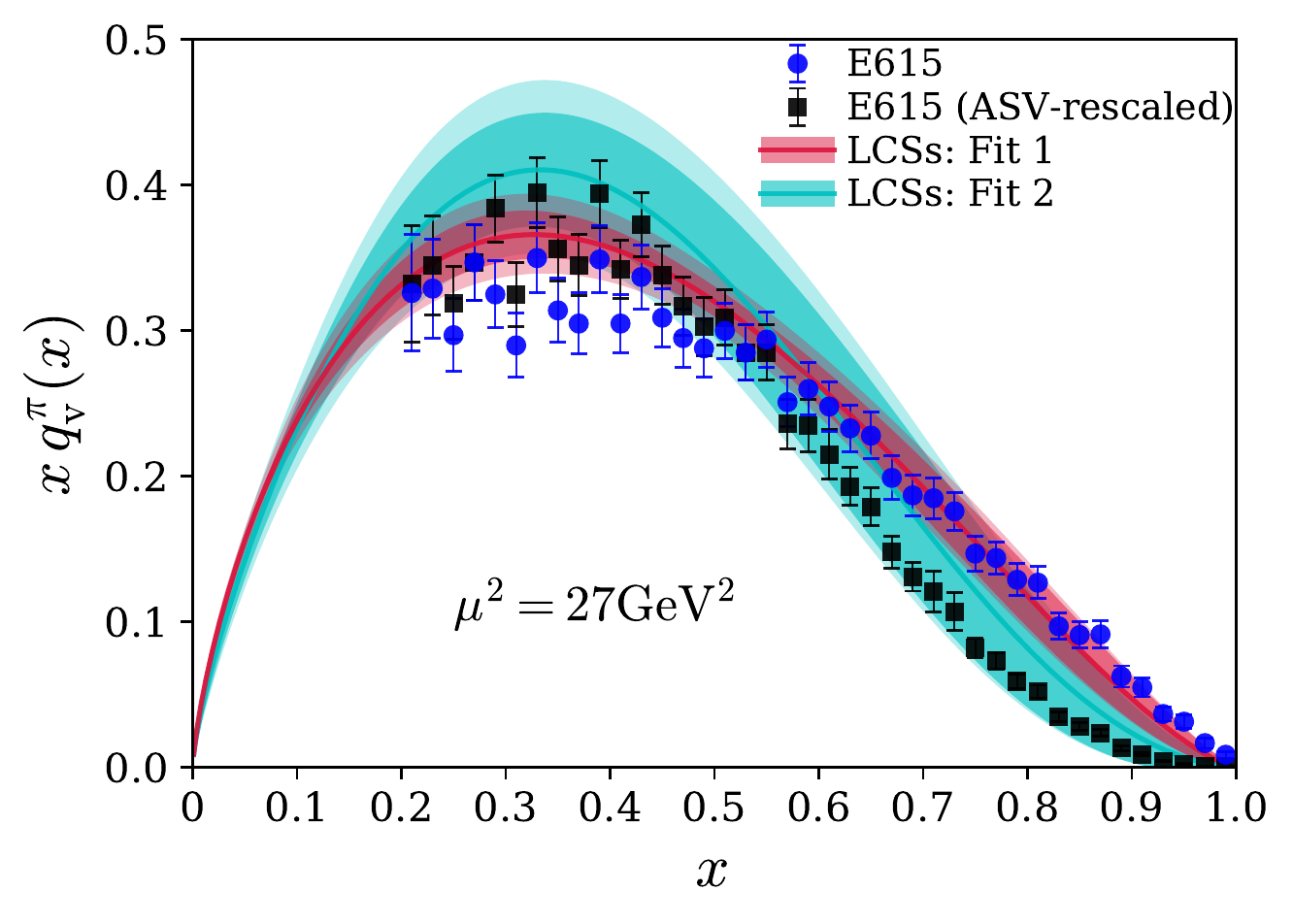}
\caption{\label{fig:evolPDF}
Comparison of pion $xq^\pi_{\rm v}(x)$ distribution obtained from this calculation with the $xq^\pi_{\rm v}(x)$ distributions extracted from the experimental Drell-Yan cross sections. The blue data points are from LO analysis~\cite{Conway:1989fs} and  the ``ASV-rescaled" black data points compiled from~\cite{Chang:2014lva} are the E615 rescaled data according to analysis~\cite{Aicher:2010cb}.  }
\end{center}
\end{figure}

For a comparison with global fits of $q^\pi_{\rm v}(x)$, we evolve our extracted PDF sets to a scale of $\mu^2=27$ GeV$^2$, from an initial scale $\mu_0=2$ GeV shown in Figure~\ref{fig:xPDF}, large enough for the validity of factorization. Fig.~\ref{fig:evolPDF} shows a comparison with the PDF extraction using LO factorization of the E615 data~\cite{Conway:1989fs}, which shows a $(1-x)$ large-$x$ behavior, and the analysis~\cite{Aicher:2010cb} where the next-to-leading-logarithmic threshold soft-gluon resummation effects~\cite{Sterman:1986aj,Catani:1989ne} are included in the calculation of the Drell-Yan cross section, which shows a softer $(1-x)^2$ falloff.   A comparison between the pion PDFs obtained from previous lattice calculations using the LCS~\cite{Sufian:2019bol}, quasi-PDFs~\cite{Izubuchi:2019lyk,Chen:2018fwa}, and pseudo-PDFs~\cite{Joo:2019bzr} methods can be found in~\cite{Joo:2019bzr}.

We need QCD factorization and perturbatively calculated matching coefficients to enable us to extract the PDFs since they are neither direct physical observables, nor directly calculable in lattice QCD.   However, QCD factorization is an approximation, and power corrections to the factorization formalism are likely more important when the observable, such as  the Drell-Yan cross section, or  the LQCD-calculated hadron matrix element, is pushed to the edge of phase space where $x\to 1$.  On the other hand, we can get some information on  the $x\to 1$ behavior of PDFs from the convolution of the factorized formalism,  by measuring  the physical cross sections or  calculating the  hadron matrix elements in LQCD not too close to the edge of phase space. However, the garnered  information on $x\sim 1$  will be mild  since the contribution from this region is much smaller than that from  the smaller $x$ regions.  This is exactly the reason why PDFs extracted from world data in QCD global analyses have a large uncertainty as $x\to 1$.  Although it might be difficult to pin down the exact ``power of $(1-x)$'' of  the pion  PDFs, the extraction of PDFs from future improved LQCD calculations of good hadron matrix elements,  the LCSs that are calculable in LQCD and factorizable to PDFs,  might help improve the accuracy of determining this ``power'', since the matching coefficients for LCSs in position space are more perturbatively stable at larger $x$ than the momentum-space matching coefficients for experimentally measured cross sections.


\section{Conclusion and Outlook}
\label{sec:6}

In this  paper we have presented the first LCS calculation of $q^\pi_{\rm v}(x)$ that incorporates results on four gauge ensembles, among these the lightest pion mass used in any lattice QCD calculation to access  $q^\pi_{\rm v}(x)$, as well as the first derivation of NLO matching coefficients from position space directly to momentum space.
The $q^\pi_{\rm v}(x)$ extracted from our LQCD calculation is remarkably consistent with that extracted from experimental data.   Given  that the NLO matching coefficient $K$ is very stable  and without large threshold logarithms  that are often seen in momentum space matching coefficients, our approach, plus future gauge ensembles with smaller lattice spacings, has the unique potential to provide a better determination of the ``power of $(1-x)$" of the $q^\pi_{\rm v}(x)$ distribution as $x\to 1$.

Central to this endeavor are  calculations in the near future with finer lattice spacings.
With a simpler nonperturbative UV renormalization, different choices of current combinations, and the nontrivial hadron-independent and stable NLO matching coefficients, the LCS  formalism  with two-current correlators is well equipped  to unravel the enigmatic structure of the pion and other hadrons, especially those that are difficult, if not impossible, to study experimentally,  complementary to other approaches, such as  the quasi- and pseudo-PDFs approaches. 
\\ 


\section*{ACKNOWLEDGMENTS}
We thank Carl  Carlson, Martha Constantinou, Luka Leskovec, Tianbo Liu, Wayne Morris, Anatoly Radyushkin, Nobuo Sato, and Jian-Hui Zhang who provided insights and expertise that greatly assisted this research. 

This work is supported by Jefferson
Science Associates, LLC under U.S. DOE Contract No. \#DE-AC05-06OR23177 within the framework of the TMD Collaboration.  
We acknowledge the facilities of the USQCD Collaboration
used for this research in part, which are funded by the Office of Science of the U.S. Department
of Energy. This material is based in part upon work supported by a grant from the Southeastern Universities Research Association (SURA) under an appropriation from the Commonwealth of Virginia. This work used the Extreme Science and Engineering Discovery Environment (XSEDE), which is supported by National Science Foundation Grant No. ACI-1548562~\cite{xsede},  and the Texas Advanced Computing Center (TACC) at the University of Texas at Austin for providing HPC resources on {\tt Frontera} that have contributed to the results in this paper.  This work was performed in part using computing facilities at William and Mary which were provided by contributions from the National Science Foundation (MRI grant PHY-1626177), the Commonwealth of Virginia Equipment Trust Fund and the Office of Naval Research. In addition, this work used resources at (National Energy Research Scientific Computing Center) NERSC, a DOE Office of Science User Facility supported by the Office of Science of the U.S. Department of Energy under Contract No. \#DE-AC02-05CH11231, as well as resources of the Oak Ridge Leadership Computing Facility at the Oak Ridge National Laboratory, which is supported by the Office of Science of the U.S. Department of Energy under Contract No. \#DE-AC05-00OR22725 (ALCC Project No. \#NPH134).  The configurations used in this work were generated under INCITE. The software codes
{\tt Chroma}~\cite{Edwards:2004sx}, {\tt QUDA}~\cite{Clark:2009wm,Babich:2010mu} and {\tt QPhiX}~\cite{QPhiX2} were used. 
The authors acknowledge support from the U.S. Department of Energy, Office of Science, Office of Advanced Scientific Computing Research and Office of Nuclear Physics, Scientific Discovery through Advanced Computing (SciDAC) program.
Also acknowledged is support from the U.S. Department of Energy Exascale Computing Project. C.E. is supported in part by the U.S. Department of Energy under Contract No. DE-FG02-04ER41302 and a Department of Energy Office of Science Graduate Student Research fellowship, through the U.S. Department of Energy, Office of Science, Office of Workforce Development for Teachers and Scientists, Office of Science Graduate Student Research (SCGSR) program. Y. M. is supported in part by the National Natural Science Foundation of China under Grants No. 11875071 and No. 11975029. K.O. acknowledges support in part  by the U.S. Department of Energy through Grant Number DE- FG02-04ER41302, by STFC consolidated Grant No. ST/P000681/1.

\providecommand{\href}[2]{#2}
\begingroup\raggedright

\endgroup


\begin{thebibliography}{10}


\bibitem{Badier:1983mj} 
  J.~Badier {\it et al.} [NA3 Collaboration],
  ``Experimental Determination of the pi Meson Structure Functions by the Drell-Yan Mechanism,''
  \href{https://doi.org/10.1007/BF01573728}{Z.\ Phys.\ C {\bf 18}, 281 (1983)}.
  
  
\bibitem{Betev:1985pf} 
  B.~Betev {\it et al.} [NA10 Collaboration],
  ``Differential Cross-section of High Mass Muon Pairs Produced by a 194-{GeV}/$c \pi^-$ Beam on a Tungsten Target,''
  \href{https://doi.org/10.1007/BF01550243}{Z.\ Phys.\ C {\bf 28}, 9 (1985)}.
  
\bibitem{Conway:1989fs} 
  J.~S.~Conway {\it et al.},
  ``Experimental Study of Muon Pairs Produced by 252-GeV Pions on Tungsten,''
  \href{https://doi.org/10.1103/PhysRevD.39.92}{Phys.\ Rev.\ D {\bf 39}, 92 (1989)}.
  
  
\bibitem{Chekanov:2002pf} 
  S.~Chekanov {\it et al.} [ZEUS Collaboration],
  ``Leading neutron production in e+ p collisions at HERA,''
  \href{https://doi.org/10.1016/S0550-3213(02)00439-X}{Nucl.\ Phys.\ B {\bf 637}, 3 (2002)}.
  
\bibitem{Aaron:2010ab} 
  F.~D.~Aaron {\it et al.} [H1 Collaboration],
  ``Measurement of Leading Neutron Production in Deep-Inelastic Scattering at HERA,''
  \href{https://doi.org/10.1140/epjc/s10052-010-1369-4}{Eur.\ Phys.\ J.\ C {\bf 68}, 381 (2010)}.
  


\bibitem{Owens:1984zj} 
  J.~F.~Owens,
  ``$Q^2$ Dependent Parametrizations of Pion Parton Distribution Functions,''
  \href{https://doi.org/10.1103/PhysRevD.30.943}{Phys.\ Rev.\ D {\bf 30}, 943 (1984)}.
  
  
  \bibitem{Aurenche:1989sx} 
  P.~Aurenche, R.~Baier, M.~Fontannaz, M.~N.~Kienzle-Focacci and M.~Werlen,
  ``The Gluon Content of the Pion From High $p_T$ Direct Photon Production,''
  \href{https://doi.org/10.1016/0370-2693(89)91351-8}{Phys.\ Lett.\ B {\bf 233}, 517 (1989)}.
  
  \bibitem{Sutton:1991ay} 
  P.~J.~Sutton, A.~D.~Martin, R.~G.~Roberts and W.~J.~Stirling,
  ``Parton distributions for the pion extracted from Drell-Yan and prompt photon experiments,''
  \href{https://doi.org/10.1103/PhysRevD.45.2349}{Phys.\ Rev.\ D {\bf 45}, 2349 (1992)}.
  
  \bibitem{Gluck:1991ey} 
  M.~Gluck, E.~Reya and A.~Vogt,
  ``Pionic parton distributions,''
 \href{https://link.springer.com/article/10.1007%2FBF01559743}{ Z.\ Phys.\ C {\bf 53}, 651 (1992)}.
 
 \bibitem{Wijesooriya:2005ir} 
  K.~Wijesooriya, P.~E.~Reimer and R.~J.~Holt,
  ``The pion parton distribution function in the valence region,''
  \href{https://doi.org/10.1103/PhysRevC.72.065203}{Phys.\ Rev.\ C {\bf 72}, 065203 (2005)}.
  
  \bibitem{Aicher:2010cb} 
  M.~Aicher, A.~Schafer and W.~Vogelsang,
  ``Soft-gluon resummation and the valence parton distribution function of the pion,''
  \href{https://doi.org/10.1103/PhysRevLett.105.252003}{Phys.\ Rev.\ Lett.\  {\bf 105}, 252003 (2010)}.
  
  \bibitem{Barry:2018ort} 
  P.~C.~Barry, N.~Sato, W.~Melnitchouk and C.~R.~Ji,
  ``First Monte Carlo Global QCD Analysis of Pion Parton Distributions,''
  \href{https://doi.org/10.1103/PhysRevLett.121.152001}{Phys.\ Rev.\ Lett.\  {\bf 121}, no. 15, 152001 (2018)}.
  




  
 \bibitem{Farrar:1979aw} 
  G.~R.~Farrar and D.~R.~Jackson,
  ``The Pion Form-Factor,''
  \href{https://doi.org/10.1103/PhysRevLett.43.246}{Phys.\ Rev.\ Lett.\  {\bf 43}, 246 (1979)}.
  
  \bibitem{Berger:1979du} 
  E.~L.~Berger and S.~J.~Brodsky,
  ``Quark Structure Functions of Mesons and the Drell-Yan Process,''
  \href{https://doi.org/10.1103/PhysRevLett.42.940}{Phys.\ Rev.\ Lett.\  {\bf 42}, 940 (1979)}.
  




  
\bibitem{Shigetani:1993dx} 
  T.~Shigetani, K.~Suzuki and H.~Toki,
  ``Pion structure function in the Nambu and Jona-Lasinio model,''
  \href{https://doi.org/10.1016/0370-2693(93)91302-4}{Phys.\ Lett.\ B {\bf 308}, 383 (1993)}.
  
  
  \bibitem{Davidson:1994uv} 
  R.~M.~Davidson and E.~Ruiz Arriola,
  ``Structure functions of pseudoscalar mesons in the SU(3) NJL model,''
  \href{https://doi.org/10.1016/0370-2693(95)00091-X}{Phys.\ Lett.\ B {\bf 348}, 163 (1995)}.
  
  

  
  
  \bibitem{Hecht:2000xa} 
  M.~B.~Hecht, C.~D.~Roberts and S.~M.~Schmidt,
  ``Valence quark distributions in the pion,''
  \href{https://doi.org/10.1103/PhysRevC.63.025213}{Phys.\ Rev.\ C {\bf 63}, 025213 (2001)}.
  
  
  \bibitem{Chen:2016sno} 
  C.~Chen, L.~Chang, C.~D.~Roberts, S.~Wan and H.~S.~Zong,
  ``Valence-quark distribution functions in the kaon and pion,"
  \href{https://doi.org/10.1103/PhysRevD.93.074021}{Phys.\ Rev.\ D {\bf 93},  074021 (2016)}.
  
    \bibitem{deTeramond:2018ecg} 
  G.~F.~de T\'eramond, T. Liu,  R.~S.~Sufian, H.~G.~Dosch, S.~J.~Brodsky and A.~Deur[HLFHS Collaboration],
  ``Universality of Generalized Parton Distributions in Light-Front Holographic QCD,''
  \href{https://doi.org/10.1103/PhysRevLett.120.182001}{Phys.\ Rev.\ Lett.\  {\bf 120}, no. 18, 182001 (2018)}.
  
  
\bibitem{Bednar:2018mtf} 
  K.~D.~Bednar, I.~C.~Clo\'et and P.~C.~Tandy,
  ``Distinguishing Quarks and Gluons in Pion and Kaon PDFs,''
  \href{https://arxiv.org/abs/1811.12310}{Phys. Rev. Lett. \textbf{124}, no.4, 042002 (2020)}.
  
\bibitem{Ding:2019lwe}
M.~Ding, K.~Raya, D.~Binosi, L.~Chang, C.~D.~Roberts and S.~M.~Schmidt,
``Symmetry, symmetry breaking, and pion parton distributions,''
\href{https://doi.org/10.1103/PhysRevLett.124.042002}{Phys. Rev. D \textbf{101}, no.5, 054014 (2020)}. 

  
    \bibitem{Jlab}
  Dasuni Adikaram, {\it et al.},  Hall A and SBS Collaboration Proposal,
  ``Measurement of Tagged Deep Inelastic Scattering (TDIS),"
  \href{https://www.jlab.org/exp_prog/proposals/15prop.html}{PR12-15-006}.
 

  
\bibitem{Denisov:2018unj} 
  B.~Adams {\it et al.},
  ``Letter of Intent: A New QCD facility at the M2 beam line of the CERN SPS (COMPASS++/AMBER),''
  \href{https://arxiv.org/abs/1808.00848}{\tt arXiv:1808.00848 [hep-ex]}.
  
\bibitem{Aguilar:2019teb} 
  A.~C.~Aguilar {\it et al.},
  ``Pion and Kaon Structure at the Electron-Ion Collider,''
  \href{https://doi.org/10.1140/epja/i2019-12885-0}{Eur.\ Phys.\ J.\ A {\bf 55}, no. 10, 190 (2019)}.
  





\bibitem{Collins:1989gx} 
  J.~C.~Collins, D.~E.~Soper and G.~F.~Sterman,
  ``Factorization of Hard Processes in QCD,''
  \href{https://doi.org/10.1142/9789814503266_0001}{Adv.\ Ser.\ Direct.\ High Energy Phys.\  {\bf 5}, 1 (1989)}.
  



  
  
  \bibitem{Harland-Lang:2014zoa} 
  L.~A.~Harland-Lang, A.~D.~Martin, P.~Motylinski and R.~S.~Thorne,
  ``Parton distributions in the LHC era: MMHT 2014 PDFs,"
  \href{https://doi.org/10.1140/epjc/s10052-015-3397-6}{Eur.\ Phys.\ J.\ C {\bf 75}, 204 (2015)}.
  
 
\bibitem{Dulat:2015mca} 
  S.~Dulat {\it et al.},
  ``New parton distribution functions from a global analysis of quantum chromodynamics,"
  \href{https://doi.org/10.1103/PhysRevD.93.033006}{Phys.\ Rev.\ D {\bf 93},  033006 (2016)}.
  
 
 \bibitem{Ball:2017nwa} 
  R.~D.~Ball {\it et al.} (NNPDF Collaboration),
  ``Parton distributions from high-precision collider data,"
  \href{https://doi.org/10.1140/epjc/s10052-017-5199-5}{Eur.\ Phys.\ J.\ C {\bf 77},  663 (2017)}.
  
  
   \bibitem{Alekhin:2017kpj} 
 S.~Alekhin, J.~Bl\"umlein, S.~Moch and R.~Pla\v{c}akyt\.e,
 ``Parton distribution functions, $\alpha_s$, and heavy-quark masses for LHC Run II,"
 \href{https://doi.org/10.1103/PhysRevD.96.014011}{Phys.\ Rev.\ D {\bf 96}, 014011 (2017)}.
  
  
\bibitem{Ethier:2017zbq} 
  J.~J.~Ethier, N.~Sato and W.~Melnitchouk,
  ``First simultaneous extraction of spin-dependent parton distributions and fragmentation functions from a global QCD analysis,''
  \href{https://doi.org/10.1103/PhysRevLett.119.132001}{Phys.\ Rev.\ Lett.\  {\bf 119}, no. 13, 132001 (2017)}.
  
  
  

  
  
  
   \bibitem{Liu:1993cv} 
  K.~F.~Liu and S.~J.~Dong,
  ``Origin of difference between $\overline d$ and $\overline u$ partons in the nucleon,"
  \href{https://doi.org/10.1103/PhysRevLett.72.1790}{Phys.\ Rev.\ Lett.\  {\bf 72}, 1790 (1994)}.
  
  
      \bibitem{Braun:2007wv} 
  V.~Braun and D.~Mueller,
  ``Exclusive processes in position space and the pion distribution amplitude,''
  \href{https://doi.org/10.1140/epjc/s10052-008-0608-4}{Eur.\ Phys.\ J.\ C {\bf 55}, 349 (2008)}.
  
  
\bibitem{Horsley:2012pz} 
  R.~Horsley {\it et al.} (QCDSF-UKQCD Collaboration),
  ``A lattice study of the glue in the nucleon,"
  \href{https://doi.org/10.1016/j.physletb.2012.07.004}{Phys.\ Lett.\ B {\bf 714}, 312 (2012)}.
    
    

  
  
\bibitem{Ji:2013dva} 
  X.~Ji,
  ``Parton physics on a Euclidean lattice,"
  \href{https://doi.org/10.1103/PhysRevLett.110.262002}{Phys.\ Rev.\ Lett.\  {\bf 110}, 262002 (2013)}.
  
  
  \bibitem{Ma:2014jla} 
  Y.~Q.~Ma and J.~W.~Qiu,
  ``Extracting Parton Distribution Functions from Lattice QCD Calculations,''
  \href{https://doi.org/10.1103/PhysRevD.98.074021}{Phys.\ Rev.\ D {\bf 98}, no. 7, 074021 (2018)}.
  
  
  \bibitem{Radyushkin:2017cyf} 
  A.~V.~Radyushkin,
  ``Quasi-parton distribution functions, momentum distributions, and pseudo-parton distribution functions,"
  \href{https://doi.org/10.1103/PhysRevD.96.034025}{Phys.\ Rev.\ D {\bf 96},  034025 (2017)}.
 




    \bibitem{Ma:2017pxb} 
  Y.~Q.~Ma and J.~W.~Qiu,
  ``Exploring Partonic Structure of Hadrons Using ab initio Lattice QCD Calculations,''
  \href{https://doi.org/10.1103/PhysRevLett.120.022003}{Phys.\ Rev.\ Lett.\  {\bf 120}, no. 2, 022003 (2018)}.
  
  
   \bibitem{Chambers:2017dov} 
  A.~J.~Chambers {\it et al.},
  ``Nucleon structure functions from operator product expansion on the lattice,"
  \href{https://doi.org/10.1103/PhysRevLett.118.242001}{Phys.\ Rev.\ Lett.\  {\bf 118}, 242001 (2017)}.
 
\bibitem{Orginos:2017kos} 
  K.~Orginos, A.~Radyushkin, J.~Karpie and S.~Zafeiropoulos,
  ``Lattice QCD exploration of parton pseudo-distribution functions,''
  \href{https://doi.org/10.1103/PhysRevD.96.094503}{Phys.\ Rev.\ D {\bf 96}, no. 9, 094503 (2017).}
  
  
    
    \bibitem{Alexandrou:2018pbm} 
  C.~Alexandrou, K.~Cichy, M.~Constantinou, K.~Jansen, A.~Scapellato and F.~Steffens,
  ``Light-Cone Parton Distribution Functions from Lattice QCD,''
  \href{https://doi.org/10.1103/PhysRevLett.121.112001}{Phys.\ Rev.\ Lett.\  {\bf 121}, no. 11, 112001 (2018)}.
  
  
  
     \bibitem{Bali:2018spj} 
  G.~S.~Bali {\it et al.},
  ``Pion distribution amplitude from Euclidean correlation functions: Exploring universality and higher-twist effects,''
  \href{https://doi.org/10.1103/PhysRevD.98.094507}{Phys.\ Rev.\ D {\bf 98}, no. 9, 094507 (2018)}.
  
  
\bibitem{Lin:2018qky} 
  H.~W.~Lin {\it et al.},
  ``Proton Isovector Helicity Distribution on the Lattice at Physical Pion Mass,''
  \href{https://doi.org/10.1103/PhysRevLett.121.242003}{Phys.\ Rev.\ Lett.\  {\bf 121}, no. 24, 242003 (2018)}.
  

  
\bibitem{Sufian:2019bol} 
  R.~S.~Sufian, J.~Karpie, C.~Egerer, K.~Orginos, J.~W.~Qiu and D.~G.~Richards,
  ``Pion Valence Quark Distribution from Matrix Element Calculated in Lattice QCD,''
  \href{https://doi.org/10.1103/PhysRevD.99.074507}{Phys.\ Rev.\ D {\bf 99}, no. 7, 074507 (2019)}.
  

  
\bibitem{Bali:2019ecy} 
  G.~S.~Bali {\it et al.} [RQCD Collaboration],
  ``Light-cone distribution amplitudes of octet baryons from lattice QCD,''
  \href{https://doi.org/10.1140/epja/i2019-12803-6}{Eur.\ Phys.\ J.\ A {\bf 55}, no. 7, 116 (2019)}.
  
\bibitem{Izubuchi:2019lyk} 
  T.~Izubuchi, L.~Jin, C.~Kallidonis, N.~Karthik, S.~Mukherjee, P.~Petreczky, C.~Shugert and S.~Syritsyn,
  ``Valence parton distribution function of pion from fine lattice,''
  \href{https://doi.org/10.1103/PhysRevD.100.034516}{Phys.\ Rev.\ D {\bf 100}, no. 3, 034516 (2019)}.
  
\bibitem{Liang:2019frk} 
  J.~Liang {\it et al.} [$\chi$QCD Collaboration],
  ``Towards the nucleon hadronic tensor from lattice QCD,''
  \href{https://doi.org/10.1103/PhysRevD.101.114503}{Phys. Rev. D \textbf{101}, no.11, 114503 (2020)}.
  
  
\bibitem{Joo:2019jct} 
  B.~Jo\'o, J.~Karpie, K.~Orginos, A.~Radyushkin, D.~Richards and S.~Zafeiropoulos,
  ``Parton Distribution Functions from Ioffe time pseudo-distributions,''
  \href{https://doi.org/10.1007/JHEP12(2019)081}{JHEP {\bf 1912}, 081 (2019)}.
  


  
 \bibitem{Cichy:2018mum} 
  K.~Cichy and M.~Constantinou,
  ``A guide to light-cone PDFs from Lattice QCD: an overview of approaches, techniques and results,''
  \href{https://doi.org/10.1155/2019/3036904}{Adv.\ High Energy Phys.\  {\bf 2019}, 3036904 (2019)}.
  
  
    
\bibitem{Yoon:2016jzj}
B.~Yoon, Y.~C.~Jang, R.~Gupta, T.~Bhattacharya, J.~Green, B.~Joó, H.~W.~Lin, K.~Orginos, D.~Richards, S.~Syritsyn and F.~Winter,
``Isovector charges of the nucleon from 2+1-flavor QCD with clover fermions,''
\href{https://doi.org/10.1103/PhysRevD.95.074508}{Phys. Rev. D \textbf{95}, no.7, 074508 (2017)}

\bibitem{Ioffe:1969kf} 
  B.~L.~Ioffe,
  ``Space-time picture of photon and neutrino scattering and electroproduction cross-section asymptotics,''
  \href{https://doi.org/10.1016/0370-2693(69)90415-8}{Phys.\ Lett.\  {\bf 30B}, 123 (1969)}.
  
\bibitem{NLOcalc}
W.~Morris, Y.~Q.~Ma, J.~W.~Qiu, A.~V.~Radyushkin, [{\emb In preparation}].

\bibitem{Sterman:1986aj} 
  G.~F.~Sterman,
  ``Summation of Large Corrections to Short Distance Hadronic Cross-Sections,''
  \href{https://doi.org/10.1016/0550-3213(87)90258-6}{Nucl.\ Phys.\ B {\bf 281}, 310 (1987)}.
  
\bibitem{Catani:1989ne} 
  S.~Catani and L.~Trentadue,
  ``Resummation of the QCD Perturbative Series for Hard Processes,''
  \href{https://doi.org/10.1016/0550-3213(89)90273-3}{Nucl.\ Phys.\ B {\bf 327}, 323 (1989)}.

\bibitem{Shimizu:2005fp}
H.~Shimizu, G.~F.~Sterman, W.~Vogelsang and H.~Yokoya,
``Dilepton production near partonic threshold in transversely polarized proton-antiproton collisions,''
\href{https://doi.org/10.1103/PhysRevD.71.114007}{Phys. Rev. D \textbf{71}, 114007 (2005)}.
 
\bibitem{Collins:1984kg}
J.~C.~Collins, D.~E.~Soper and G.~F.~Sterman,
``Transverse Momentum Distribution in Drell-Yan Pair and W and Z Boson Production,''
\href{https://doi.org/10.1016/0550-3213(85)90479-1}{Nucl. Phys. B \textbf{250}, 199-224 (1985)}

\bibitem{Li:2020xml}
Z.~Y.~Li, Y.~Q.~Ma and J.~W.~Qiu,
``Extraction of Next-to-Next-to-Leading-Order PDFs from Lattice QCD Calculations,''
[\href{https://arxiv.org/abs/2006.12370}{\tt arXiv:2006.12370 [hep-ph]}].

\bibitem{lattices}
R.~Edwards, B.~Jo\'o, K.~Orginos, D.~Richards, and F.~Winter
``U.S. 2+1 flavor clover lattice generation program,"
Unpublished (2016).

  
\bibitem{Boyd:1994tt} 
  C.~G.~Boyd, B.~Grinstein and R.~F.~Lebed,
  ``Constraints on form-factors for exclusive semileptonic heavy to light meson decays,''
  \href{https://doi.org/10.1103/PhysRevLett.74.4603}{Phys.\ Rev.\ Lett.\  {\bf 74}, 4603 (1995)}.
  
  \bibitem{Bourrely:2008za} 
  C.~Bourrely, I.~Caprini and L.~Lellouch,
  ``Model-independent description of $B \to \pi l \nu$ decays and a determination of $|V(ub)|$,''
  \href{https://doi.org/10.1103/PhysRevD.82.099902}{Phys.\ Rev.\ D {\bf 79}, 013008 (2009)}
  \href{https://doi.org/10.1103/PhysRevD.79.013008}{Erratum: [Phys.\ Rev.\ D {\bf 82}, 099902 (2010)]}.

  \bibitem{Briceno:2018lfj} 
  R. A. Brice\~no, J.~V.~Guerrero, M.~T.~Hansen and C.~J.~Monahan,
  ``Finite-volume effects due to spatially nonlocal operators,''
  \href{https://doi.org/10.1103/PhysRevD.98.014511}{Phys.\ Rev.\ D {\bf 98}, no. 1, 014511 (2018)}.

\bibitem{Joo:2019bzr} 
  B.~Jo\'o, J.~Karpie, K.~Orginos, A.~V.~Radyushkin, D.~G.~Richards, R.~S.~Sufian and S.~Zafeiropoulos,
  ``Pion Valence Structure from Ioffe Time Pseudo-Distributions,''
  \href{https://doi.org/10.1103/PhysRevD.100.114512}{Phys.\ Rev.\ D {\bf 99}, no. 11, 114512 (2019)}.
  
\bibitem{Gribov:1972ri} 
  V.~N.~Gribov and L.~N.~Lipatov,
  ``Deep inelastic e p scattering in perturbation theory,''
  Sov.\ J.\ Nucl.\ Phys.\  {\bf 15}, 438 (1972)
  [Yad.\ Fiz.\  {\bf 15}, 781 (1972)].
  
\bibitem{Altarelli:1977zs} 
  G.~Altarelli and G.~Parisi,
  ``Asymptotic Freedom in Parton Language,''
  \href{https://doi.org/10.1016/0550-3213(77)90384-4}{Nucl.\ Phys.\ B {\bf 126}, 298 (1977)}.
  
\bibitem{Dokshitzer:1977sg} 
  Y.~L.~Dokshitzer,
  ``Calculation of the Structure Functions for Deep Inelastic Scattering and e+ e- Annihilation by Perturbation Theory in Quantum Chromodynamics.,''
  Sov.\ Phys.\ JETP {\bf 46}, 641 (1977)
  [Zh.\ Eksp.\ Teor.\ Fiz.\  {\bf 73}, 1216 (1977)].

\bibitem{Regge:1959mz}
T.~Regge,
``Introduction to complex orbital momenta,''
\href{https://doi.org/10.1007/BF02728177}{Nuovo Cim. \textbf{14}, 951 (1959)}. 
  
\bibitem{Brodsky:1973kr}
S.~J.~Brodsky and G.~R.~Farrar,
``Scaling Laws at Large Transverse Momentum,''
\href{https://doi.org/10.1103/PhysRevLett.31.1153}{Phys. Rev. Lett. \textbf{31}, 1153-1156 (1973)}

\bibitem{ROOT} 
  Rene Brun and Fons Rademakers,
``ROOT - An Object Oriented Data Analysis Framework,"
\href{https://root.cern.ch/publications}{Proceedings AIHENP'96 Workshop, Lausanne, Sep. 1996, Nucl. Inst. \& Meth. in Phys. Res. A 389 (1997) 81-86}.
 
\bibitem{Buckley:2014ana} 
  A.~Buckley, J.~Ferrando, S.~Lloyd, K.~Nordstr\"om, B.~Page, M.~R\"ufenacht, M.~Sch\"onherr and G.~Watt,
  ``LHAPDF6: parton density access in the LHC precision era,''
  \href{https://doi.org/10.1140/epjc/s10052-015-3318-8}{Eur.\ Phys.\ J.\ C {\bf 75}, 132 (2015)}.
  
\bibitem{Karpie:2019eiq}
J.~Karpie, K.~Orginos, A.~Rothkopf and S.~Zafeiropoulos,
``Reconstructing parton distribution functions from Ioffe time data: from Bayesian methods to Neural Networks,''
\href{https://doi.org/10.1007/JHEP04(2019)057}{JHEP \textbf{04}, 057 (2019)}.

\bibitem{Chang:2014lva} 
  L.~Chang, C.~Mezrag, H.~Moutarde, C.~D.~Roberts, J.~RodrÃ­guez-Quintero and P.~C.~Tandy,
  ``Basic features of the pion valence-quark distribution function,''
  \href{https://doi.org/10.1016/j.physletb.2014.08.009}{Phys.\ Lett.\ B {\bf 737}, 23 (2014)}.
  

 
\bibitem{Chen:2018fwa} 
  J.~H.~Zhang, J.~W.~Chen, L.~Jin, H.~W.~Lin, A.~Sch\"afer and Y.~Zhao,
  ``First direct lattice-QCD calculation of the $x$-dependence of the pion parton distribution function,''
  \href{https://doi.org/10.1103/PhysRevD.100.034505}{Phys.\ Rev.\ D {\bf 100}, 034505 (2019)}.


\bibitem{xsede}
J. Towns et al., ``XSEDE: Accelerating Scientific Discovery,"  
\href{http://ieeexplore.ieee.org/stamp/stamp.jsp?tp=&arnumber=6866038&isnumber=6924616}{Computing in Science \& Engineering, vol. 16, no. 5, pp. 62-74, Sept.-Oct. 2014}.


\bibitem{Edwards:2004sx} 
  R.~G.~Edwards {\it et al.} [SciDAC and LHPC and UKQCD Collaborations],
  ``The Chroma software system for lattice QCD,''
\href{https://doi.org/10.1016/j.nuclphysbps.2004.11.254}{Nucl.\ Phys.\ Proc.\ Suppl.\  {\bf 140}, 832 (2005)}.


\bibitem{Clark:2009wm} 
  M.~A.~Clark, R.~Babich, K.~Barros, R.~C.~Brower and C.~Rebbi,
  ``Solving Lattice QCD systems of equations using mixed precision solvers on GPUs,''
\href{https://doi.org/10.1016/j.cpc.2010.05.002}{Comput.\ Phys.\ Commun.\  {\bf 181}, 1517 (2010)}.

\bibitem{Babich:2010mu} 
  R.~Babich, M.~A.~Clark and B.~Joo,
  ``Parallelizing the QUDA Library for Multi-GPU Calculations in Lattice Quantum Chromodynamics,''
\href{https://arxiv.org/abs/1011.0024}{\tt arXiv:1011.0024 [hep-lat]}.

\bibitem{QPhiX2}
B.~Jo\'o, D.~D.~Kalamkar, T.~Kurth, K.~Vaidyanathan, A.~Walden,
``Optimizing Wilson-Dirac Operator and Linear Solvers for Intel® KNL"
\href{https://doi.org/10.1007/978-3-319-46079-6_30}{2016. ISC: High Performance Computing, pp 415-427 30.}

\end{thebibliography}
\end{document}